\documentclass[aps,prc,showpacs,superscriptaddress]{revtex4-1}
\usepackage{graphicx}
\usepackage{amsmath,amssymb}
\usepackage{hyperref}
\usepackage{color}
\usepackage{multirow}
\usepackage{ulem}
\usepackage[version=3]{mhchem}

\begin{document}

\title{Deep learning for nuclear masses in deformed relativistic Hartree-Bogoliubov theory in continuum}

\author{Soonchul Choi}
\affiliation{Center for Exotic Nuclear Studies, Institute for Basic Science, Daejeon 34126, Korea}

\author{Kyungil Kim}
\affiliation{Institute for Rare Isotope Science, Institute for Basic Science, Daejeon 34000, Korea}

\author{Zhenyu He}
\affiliation{School of Physics, International Research Center for Big-Bang Cosmology and Element Genesis, Peng Huanwu Collaborative Center for Research and Education, Beihang University, Beijing 100191, China}

\author{Youngman~Kim}
\email{ykim@ibs.re.kr}
\affiliation{Center for Exotic Nuclear Studies, Institute for Basic Science, Daejeon 34126, Korea}
\affiliation{School of Physics, International Research Center for Big-Bang Cosmology and Element Genesis, Peng Huanwu Collaborative Center for Research and Education, Beihang University, Beijing 100191, China}

\author{Toshitaka Kajino}
\affiliation{School of Physics, International Research Center for Big-Bang Cosmology and Element Genesis, Peng Huanwu Collaborative Center for Research and Education, Beihang University, Beijing 100191, China}
\affiliation{National Astronomical Observatory of Japan, Mitaka, Tokyo, 181-8588, Japan}
\affiliation{Graduate School of Science, The University of Tokyo, Hongo, Tokyo, 113-033, Japan}

\date{\today}

\begin{abstract}
Most nuclei are deformed, and these deformations play an important role in various nuclear and astrophysical phenomena. Microscopic nuclear mass models have been developed based on covariant density functional theory to explore exotic nuclear properties. Among these, we adopt mass models based on the relativistic continuum Hartree-Bogoliubov theory (RCHB) with spherical symmetry and the deformed relativistic Hartree-Bogoliubov theory in continuum (DRHBc) with axial symmetry to study the effects of deformation on the abundances produced during the rapid neutron-capture process (r-process).

Since the DRHBc mass table has so far been completed only for even-Z nuclei, we first investigate whether a Deep Neural Network (DNN) can be used to extend the DRHBc mass table by focusing on nuclear binding energies. To incorporate information about odd-odd and odd-even isotopes into the DNN, we also use binding energies from AME2020 as a training set, in addition to those from the DRHBc mass table for even-Z nuclei. After generating an improved mass table through the DNN study, we conduct a sensitivity analysis of r-process abundances to deformation or mass variations using the RCHB$^\star$ and DRHBc$^\star$ mass tables (where $\star$ indicates that the mass table is obtained from the DNN study).
For the r-process sensitivity study, we consider magnetohydrodynamic jets and collapsar jets. Our findings indicate that r-process abundances are sensitive to nuclear deformation, particularly within the mass range of $A=80-120$.
\end{abstract}

\pacs{}

\maketitle

\section{Introduction}
The mass of a nucleus is one of the fundamental nuclear properties, and as is well known it is not just the sum of  its constituent nucleon masses. 
The nuclear mass is a basic observable to dissect theoretically internal structures of a nucleus such as deformations and paring gaps.  It is also essential to determine the Q-value for nuclear reactions in astrophysical environments and so important to understand the origin of the elements~\cite{Lunney:2003zz}.
The production of elements up to iron, as well as about half of the elements heavier than iron, is reasonably well explained by processes such as Big Bang nucleosynthesis and slow neutron capture. The  other half of the heavy elements in the universe is attributed to the r-process. 
For a recent review of the r-process, we refer to Refs.~\cite{Bertulani:2016eru, Kajino:2019abv, Horowitz:2018ndv,Cowan:2019pkx}.
The sensitivity of the r-process nucleosynthesis to nuclear masses has been extensively studied, see Refs.~\cite{Mumpower:2015ova, Kajino:2016pia} for a review and Refs.~\cite{Mumpower:2015hva, Jiang:2021pvh, Hao:2023bzz} for some recent works.

Many efforts have been made to theoretically develop a reliable global nuclear mass model: e.g.
finite-range droplet model (FRDM)~\cite{Moller:1993ed},  Weiz\"sacker-Skyrme (WS) model~\cite{Wang:2010uk}
and Hartree-Fock-Bogoliubov method based on non-relativistic density functional theory~
\cite{Goriely:2009zzb, Goriely:2009zz, Goriely:2010bm}.

Relativistic density functional theory has also been developed for nuclear properties with much success.
To accurately capture the characteristics of loosely bound nuclei, it is crucial to treat pairing correlations and continuum effects in a self-consistent manner, as the Fermi energy of such nuclei is typically close to the continuum threshold.
This feature was successfully implemented in the relativistic continuum Hartree-Bogoliubov (RCHB) theory~\cite{Meng:1996zz,Meng:1998axq, Meng:2005jv, Vretenar:2005zz, Xia:2017zka} with the point coupling density functional, PC-PK1~\cite{Zhao:2010hi}, where the Dirac equations are solved by the shooting method in the coordinate space.

Nuclear deformations are common across most nuclei, particularly in exotic ones. To accurately describe deformed nuclei, the RCHB theory was extended to the deformed relativistic Hartree-Bogoliubov theory in the continuum (DRHBc) with meson-exchange interactions~\cite{Zhou:2009sp, Li:2012gv}, where  the deformed relativistic Hartree-Bogoliubov equations are solved in a Dirac Woods-Saxon basis. 
DRHBc with contact interactions was developed, and its application to even-even nuclei was discussed in detail in Ref.~\cite{DRHBcMassTable:2020cfw}; recently, it is extended further to odd-A and odd-odd nuclei~\cite{DRHBcMassTable:2022rvn}. 
A difference in the parameters between DRHBc and  RCHB is the paring strength. The pairing strength in DRHBc is$-325.0$ MeV fm$^3$ and that in RCHB is $-342.5$ MeV fm$^3$.
DRHBc  with PC-PK1 has been applied to investigate various exotic nuclear properties~\cite{Pan:2021oyq, An:2021yoj, Kim:2021skf, Choi:2022rdj, Sun:2022gdu, Zhang:2023dhj, Zhang:2023bqg, Xiao:2023uld, Zhang:2023fym} and mass tables for even Z - even N~\cite{DRHBcMassTable:2022uhi} and for even-odd~\cite{DRHBcMassTable:2024nvk}. Here "Z" and "N" denote the proton and neutron numbers in a nucleus. As summarized in Table 1 of  Ref.~\cite{DRHBcMassTable:2024nvk}, for the even-Z nuclei the DRHBc mass table achieved the root-mean-square deviation (RMS) $1.433$ MeV with respect to the AME2020 data~\cite{Wang:2021xhn} among nonrelativistic and relativistic microscopic density functional calculations; for example the nonrelativistic  Hartree-Fock-Bogoliubov (HFB) theory with the UNEDF1 density functional obtained the rms deviation of $1.934$ MeV which is the smallest
rms deviation among the HFB theory with various density functionals.
Since the main difference between RCHB and DRHBc is deformations which lead to different mass predictions for deformed nuclei, it will be interesting to investigate the sensitivity of the r-process abundances to nuclear deformations or masses. However, so far the DRHBc mass table is  completed only for even-even and even-odd nuclei.

In this work, we first predict the masses of the odd-odd and odd-even nuclei using a deep neural network (DNN).
For a recent study on the machine learning applied to the nuclear masses we refer to Refs.~\cite{Niu:2018csp, Negoita:2018kgi, Jiang:2019zkg, Li:2022ifg, Zeng:2022azv, Wu:2022nnc, Niu:2022gwo, Mumpower:2022peg, Knoll:2022abg, Mumpower:2023lch, Yuksel:2024zky,  Li:2023dsv} and Refs.~\cite{Boehnlein:2021eym, Garcia-Ramos:2023rtd} for a recent review.
To validate the performance of the DNN in the present study, we first use the RCHB data and AME2020~\cite{Wang:2021xhn} as a training data set. When we use the RCHB data, we exclude odd-Z data to see if  the neural network system can learn about odd-Z information from the AME2020 data.  After training, we compare the predicted data with  AME2020  and also with RCHB combined with AME2020 data sets and obtain  the RMS deviations of $1.599$ MeV and $2.457$ MeV, respectively. 
When the original RCHB mass table was compared with AME2020, the RMS deviation  was 7.980 MeV. Though we have used AME2020 as a training set, it is a substantial improvement. We then apply the same method to the DRHBc case. 

Finally, we calculate the r-process abundances by using some part of the RCHB$^\star$ and DRHBc$^\star$ mass tables, where the mass difference exceeds 5 MeV, in astrophysical sites of the r-process: magnetohydrodynamic (MHD) jets and collapsars.

\section{Neural Network Model}
In this study, we use a deep neural network (DNN) consisting of the first hidden layer, intermediate hidden layer, last hidden layer and  output layer. The relations between the inputs and each layer are as follows:

With the first hidden layer, the relation is given by
\begin{equation}
  a_j^{(1)} = E(\sum_{i=1}^{n}w_{ij}^{(1)} x_i + b_j^{(1)})\, ,
\end{equation}
with the intermediate hidden layer $l=2,3,...,j-1$
\begin{equation}
  a_j^l = E(\sum_{i=1}^{k}w_{ij}^l a_i^{l-1} + b_j^{l})\, ,
\end{equation}
with the last hidden layer
\begin{equation}
  a_j^j = E(\sum_{i=1}^{k}w_{ij}^j a_i^{j-1} + b_j^j)
\end{equation}
and with the output layer
\begin{equation}
  z^{j+1}_\textmd{out} = \sum_{i=1}^{k}w_{i\textmd{out}}^{j+1}a_i^j + b_\textmd{out}^{j+1}\, .
\end{equation}
Here $x_i, w_{ij}^{l}, b_j^l, a_j^l$, and $z_\textmd{out}$ are the input data, the weight from a node $i$ in a layer $l-1$ to a node $j$ in a layer $l$, biases for a node $j$ in a layer $l$, outputs of a node $j$ in a layer $l$ and output of the output layer, respectively.
  In the above relations, $E$ is the Elu function employed as the activation function~\cite{Tensorflow},
\begin{eqnarray}
  &&E(x)= x\quad x \ge 0,
  \nonumber\\
  &&E(x)= \alpha(e^x-1) \quad x < 0,
\end{eqnarray}
where $\alpha = 1$. 
 
 The inputs of the neutral network are the proton number ($Z$), neutron number ($N$), nuclear pairing ($\delta$) and shell effect ($P$).
$\delta$ is defined by $\delta=[(-1)^Z+(-1)^N]/2$, and $P$ is given by
\begin{equation}
P=\frac{\nu_p\nu_n}{\nu_p+\nu_n}\, ,
\end{equation}
Here, $\nu_p$ and $\nu_n$ are the differences between the actual nucleon numbers $Z$ and $N$ and the nearest magic numbers~\cite{Kirson:2008yvv}. The value of $\nu_p$ and $\nu_n$ is zero at a closed shell and reaches a maximum at the mid-shell.
When one considers the shell effect differently for protons and neutrons, one can use $P_n=\nu_n$ and $P_p=\nu_p$~\cite{Mumpower:2023lch}.

In this study we consider three cases with different inputs: two-inputs ($Z,N$), four-inputs ($Z,N,\delta,P$), five-inputs ($Z,N,\delta,P_n,P_p$). We remark here that we also investigated the case with different pairing parameters for the proton and neutron
and found that the results do not improve.
We use the root-mean-square (RMS) as the objective function $\sigma$,
\begin{equation}
\sigma = \sqrt{\sum_i^n \frac{(E_i^{\rm{trained}}-E_i^{\rm{ref}})^2}{n}}\, ,
\end{equation}
where $n$ is the number of data. Here $E_i^{\rm{ref}}$ denotes the binding available binding energies from AME2020, DRHBc and RCHB. 

To obtain the odd-even and odd-odd binding energies using a deep neural network, the binding energies of even-even and even-odd isotopes from DRHBc calculations are used as a training set. In addition, to include information about odd-odd and odd-even isotopes to the deep neural network system we also use the binding energies in AME2020~\cite{Wang:2021xhn} as a training set. In case the binding energy of an isotope
is available in both DRHBc and AME2020, we take the value from AME2020.
In a DNN study, in general one divides the available data into training, validation and test data sets.
In this work since we have no available data for odd-odd and odd-even nuclei in the DRHBc results, we perform our DNN study as follows.
We first consider the RCHB and AME2020 data. Though the binding energies of the odd-odd and odd-even isotopes are available in RCHB, 
we use the AME2020 data and  even-Z results of RCHB to train the DNN system and predict the whole mass table from Z=8 to 120. 
We then evaluate the RMS deviations to judge if our DNN system predicts the binding energies reliably. 
If this procedure works, we do the same study with DRHBc.

Now, we use the RCHB data and AME2020~\cite{Wang:2021xhn} as a training data set. 
When we employ the RCHB data, we exclude odd-Z data to see if  the neural network systems can learn about odd-Z information from the AME2020 data. 
 After training with the five inputs, the predicted data are compared with both AME2020 and RCHB combined with AME2020 data sets and the corresponding  RMS deviations are $1.862$ MeV and $1.957$ MeV, respectively. 
When the original RCHB mass table was compared with AME2020, the RMS deviation was 7.980 MeV. Though we use AME2020 as a training set, it is a substantial improvement. This successful method is then applied to the DRHBc case.

\section{DNN results}
Before we perform a sample mass sensitivity study in r-abundances, we present our results from the DNN study. 

In Table~\ref{tab1}, we show the RMS deviation of RCHB$^\star$ and DRHBc$^\star$ compared with the AME2020 data sets, where
it can be seen  that the case with the five-inputs results in the smallest RMS deviation (0.842 MeV) for DRHBc$^\star$. 
 For comparison, the RMS deviation between AME2020 and the even-Z DRHBc$^\star$ mass table is 1.433 MeV.
Therefore, we present our results with the five-inputs. 
We remark here again that to obtain RCHB$^\star$ we don't use odd-Z information from RCHB.
\begin{table*}
  \begin{center}
  \begin{tabular}{c|cccc}\hline 
           \phantom{aa}& \phantom{aa} Two-inputs   \phantom{aa} & \phantom{aa} Four-inputs   \phantom{aa} & \phantom{aa} Five-inputs   \phantom{aa} \\
           \hline
           RMS deviation (RCHB$^\star$) & $1.779$  & $1.584$ & $1.862$ \\
           RMS deviation (DRHBc$^\star$) & $1.747$  & $1.541$ & $0.842$ \\
           \hline
  \end{tabular}
  \end{center}
  \caption{The RMS deviations between AME2020~\cite{Wang:2021xhn} and predicted masses in units of MeV.}
  \label{tab1}
\end{table*}
 
In Table \ref{tab2}, we compare the masses of the newly measured neutron-rich isotopes in Ref.~\cite{Xian:2024ixi}, 
where the masses of $^{88,89}$As were measured for the first time,  with our results in DRHBc$^\star$ and obtain the RMS deviation of $0.725$ MeV.
This deviation is a bit improved compared with $0.842$~MeV in Table~\ref{tab1}, which indicates that  DRHBc$^\star$ (and also  DRHBc) may work better for
exotic nuclei.
\begin{table*}
  \begin{center}
  \begin{tabular}{|c|cc||c|cc|}\hline
  & Exp & DRHBc$^\star$ & & Exp & {DRHBc}$^\star$  \\
  \hline 
$^{82}$Ge	 &	-65413.7	& -66241.4 & $^{86}$Br	& -75638.4 & -75753.7 \\
$^{82}$As	 &	-70107.2	& -71057.0 & $^{87}$As	& -55617.8 & -56543.6 \\
$^{82}$Se	 &	-77589.8  &	-78087.6 & $^{87}$Se	&	-66425.3 & -66893.0 \\
$^{88}$Se	 &	-63882.5	& -64558.6 & $^{88}$As  &	-50677.8 & -51316.9 \\
$^{83}$Ga	 &	-49260.8  &	-49916.4 & $^{89}$As	&	-46686.5 & -48107.6 \\
$^{83}$Ge	 &	-60975.6  &	-61449.0 & $^{89}$Se	&	-58989.4 & -59523.4 \\
$^{83}$As	 &	-69670.0  & -70609.5 & $^{89}$Br  &	-68275.0 & -69105.2 \\ 
$^{84}$Ga	 &	-44088.5  &	-44293.8 & $^{89}$Kr	&	-76537.0 & -76086.8 \\ 
$^{84}$Ge	 &	-58148.0  & -58600.7 & $^{90}$Se  &	-55881.3 & -56777.5 \\  
$^{84}$As	 &	-65853.5  &	-66646.1 & $^{90}$Br	&	-63998.8 & -64679.2 \\ 
$^{84}$Se	 &	-75952.5  &	-76627.1 & $^{90}$Rb	&	-79352.3 & -79180.9 \\ 
$^{85}$Ge	 &	-53116.2  &	-53231.4 & $^{91}$Se	&	-50267.4 & -51382.0 \\ 
$^{85}$As	 &	-63196.4  &	-64369.0 & $^{91}$Br	&	-61106.8 & -62240.5 \\ 
$^{85}$Se	 &	-72415.9  &	-72619.9 & $^{91}$Kr	&	-70971.1 & -71101.7 \\ 
$^{85}$Br	 &	-78599.0  & -79468.2 & $^{91}$Rb  &	-77760.6 & -77942.6 \\
$^{86}$Ge	 &	-49596.7  &	-49951.6 & $^{92}$Br  &	-56240.5 & -57256.0 \\ 
$^{86}$As	 &	-58964.6  &	-59529.7 & $^{92}$Kr	&	-68772.0 & -69465.0 \\
$^{86}$Se	 &	-70503.5  &	-71228.0 &  &  & \\ 
\hline 
\end{tabular}
  \end{center}
  \caption{ The mass excess of the newly measured isotopes~\cite{Xian:2024ixi} compared with that of the calculated ones in DRHBc$^\star$ in units of keV.}
  \label{tab2}
\end{table*}

\begin{figure}[ht!] 
  \begin{center}
  \includegraphics[width=6cm]{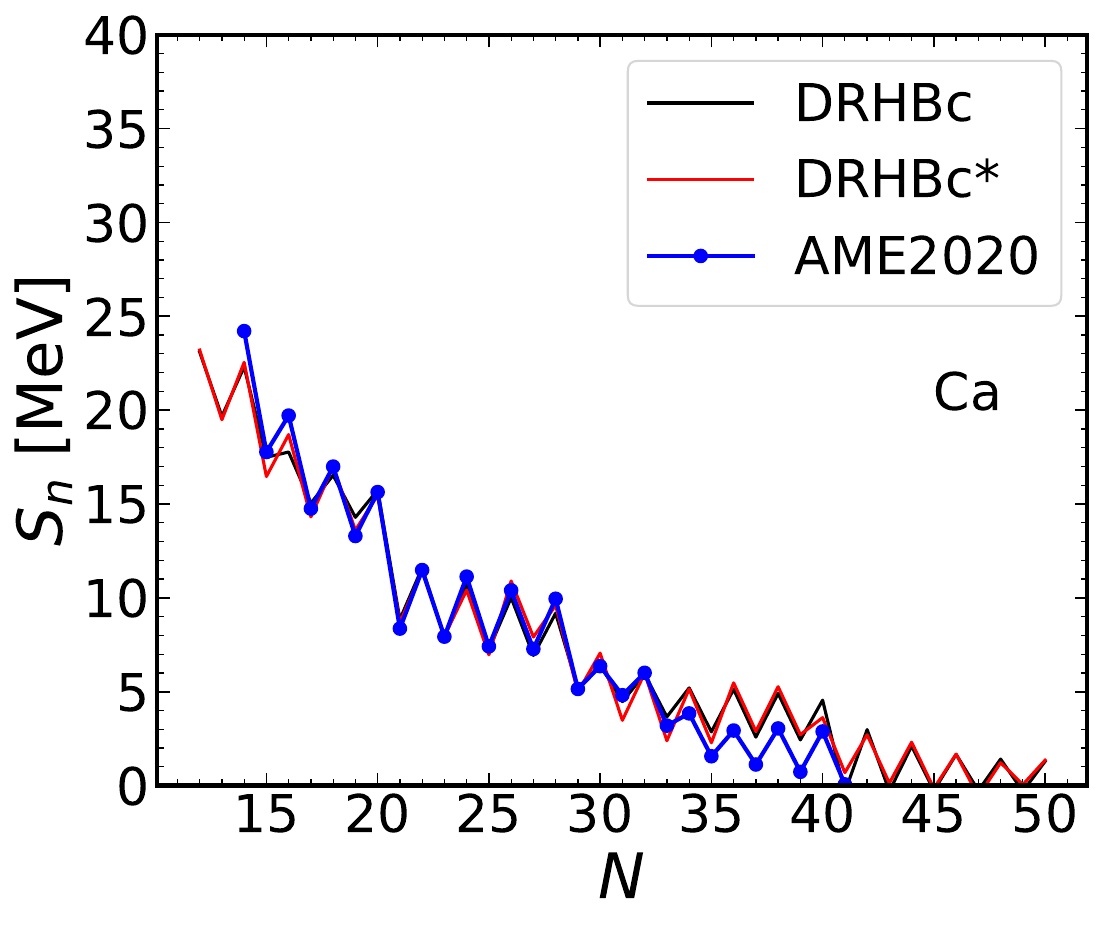}
     \includegraphics[width=6cm]{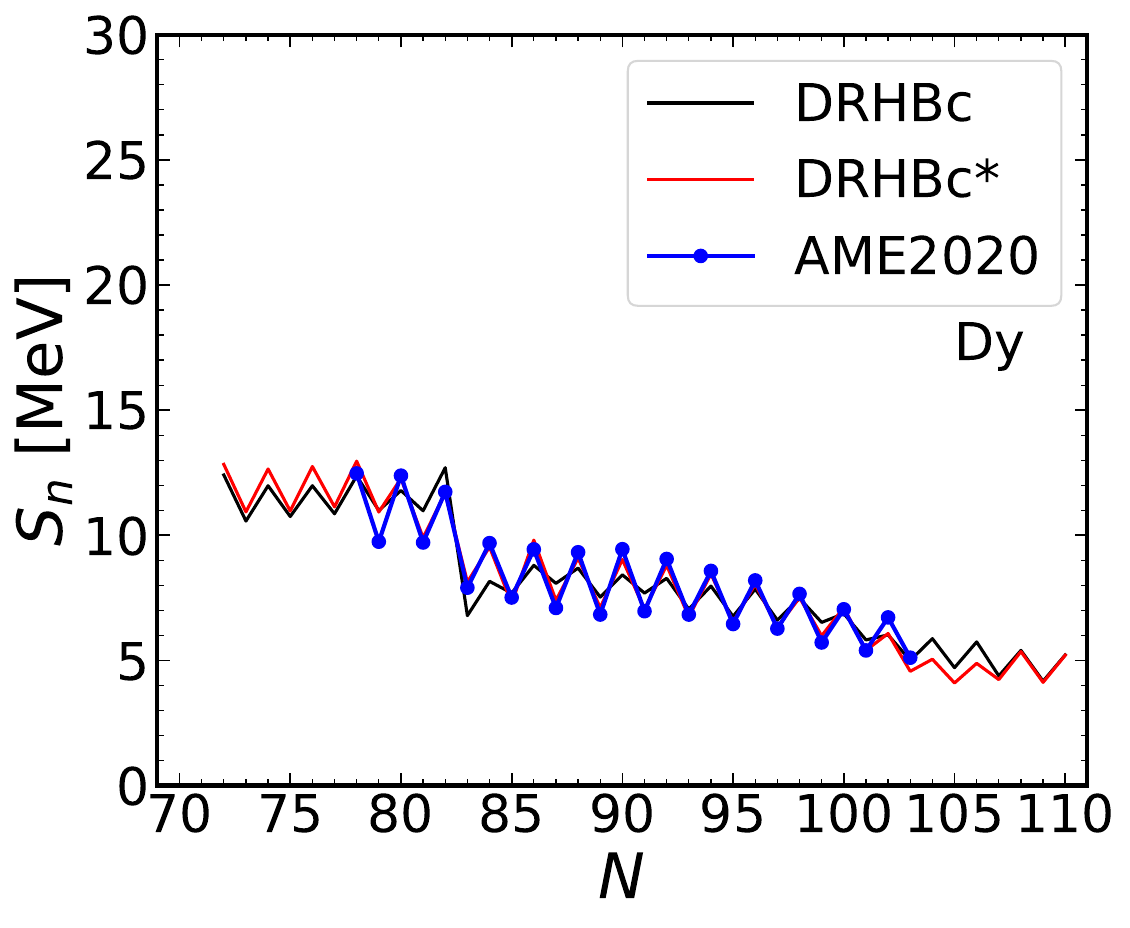}
    \includegraphics[width=6cm]{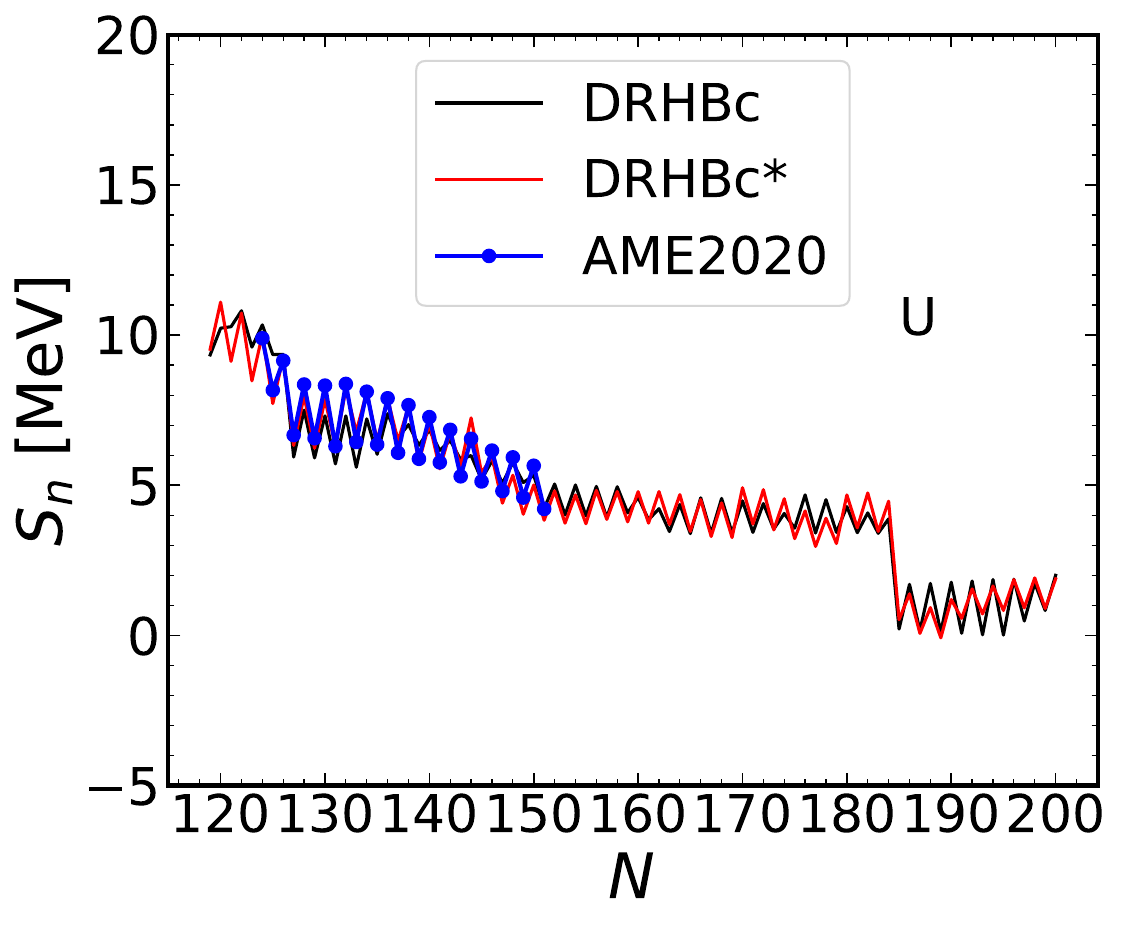}
    \includegraphics[width=6cm]{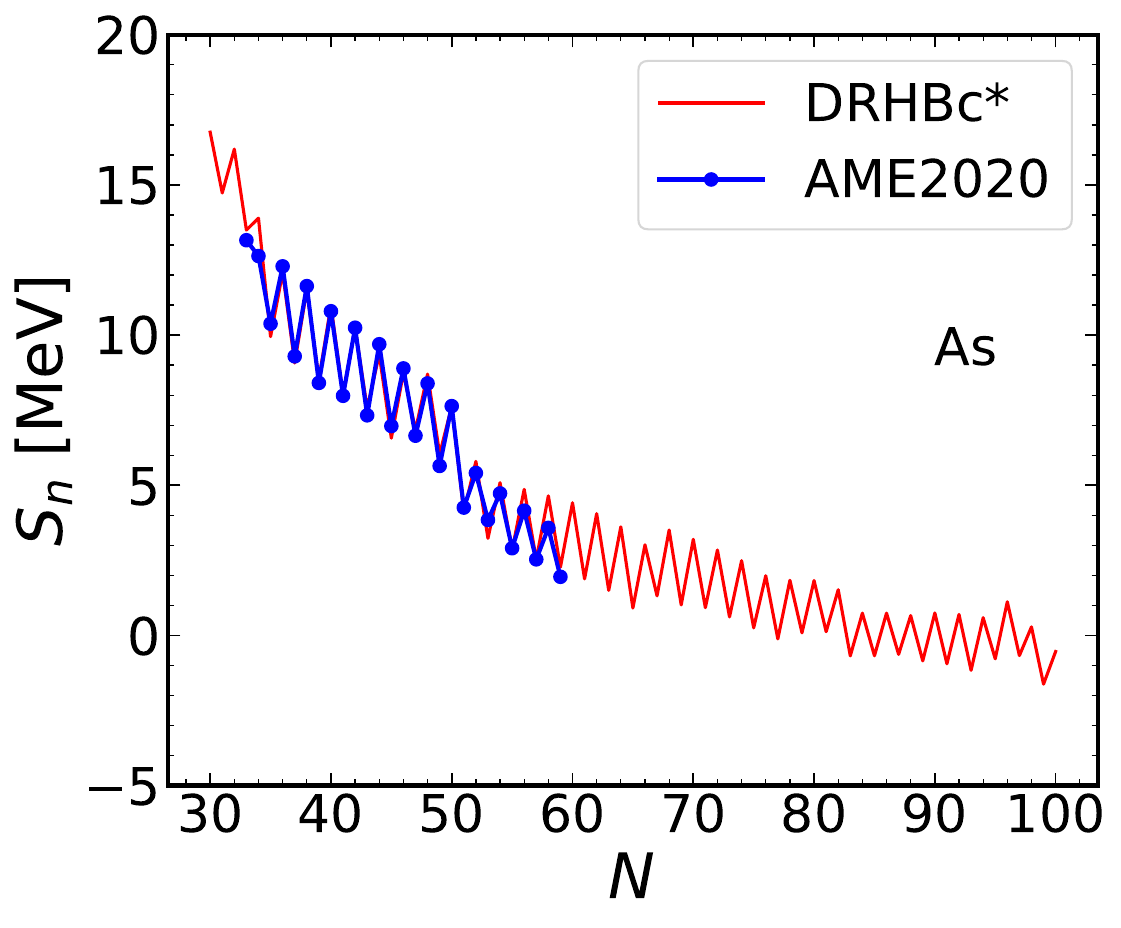}
  \end{center}
  \caption{ One neutron separation energies of the Ca, Dy, U and As isotopes from DRHBc and DRHBc$^\star$ compared with those from AME2020~\cite{Wang:2021xhn}. Note that for the As isotopes we compare the separation energies only with  AME2020, as DRHBc is not yet available for odd-Z nuclei.}
   \label{Sn}
  \end{figure}
Figure~\ref{Sn} shows the one-neutron separation energy $\rm S_n$ of a few selected isotopes from DRHBc and DRHBc$^\star$
compared with the experiments. Our results exhibit, as expected, an odd-even staggering. It can be seen from Fig.~\ref{Sn}
that our results from DRHBc$^\star$ agree with the available experimental data.

\begin{figure}[ht!]
  \begin{center}
  \includegraphics[width=13cm]{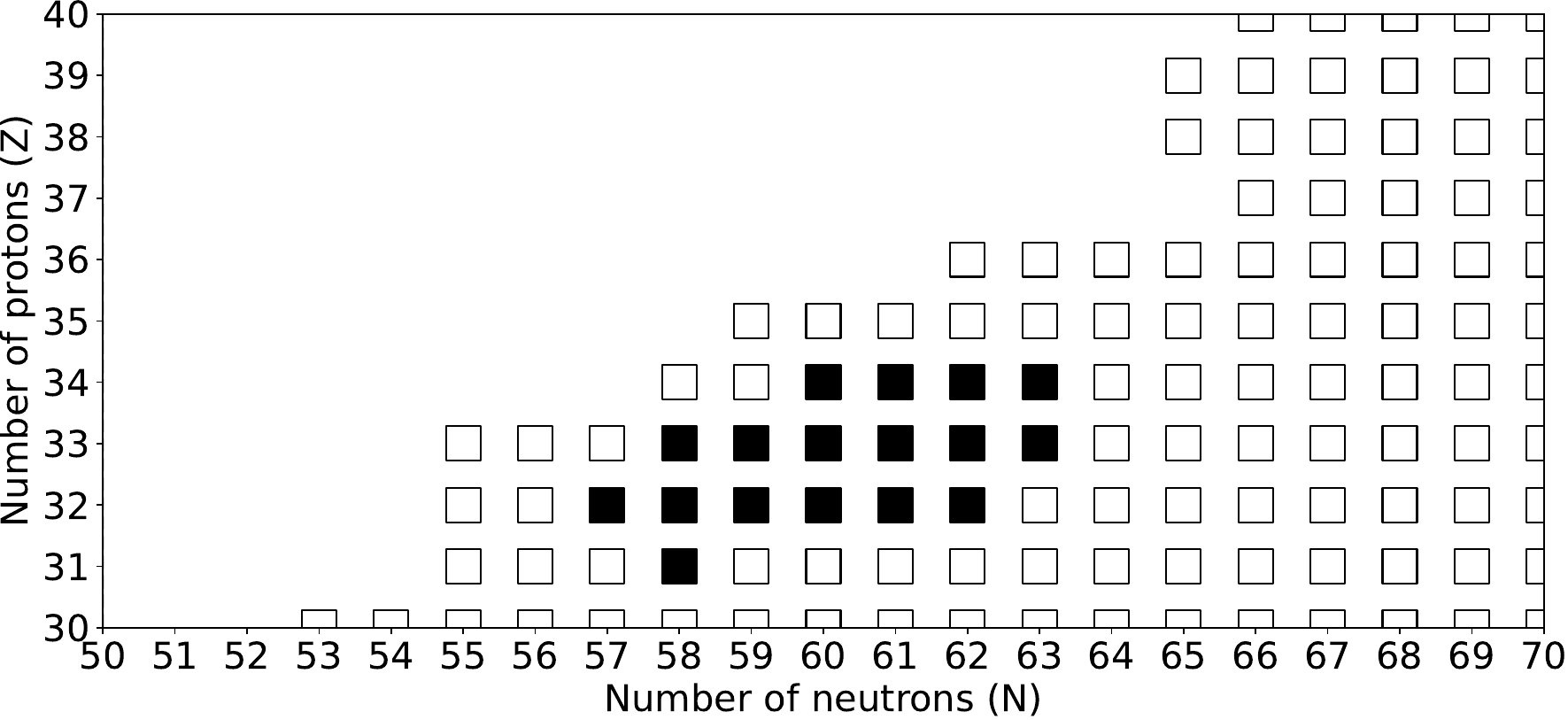}
  \end{center}
  \caption{The isotopes with a mass difference greater than 5 MeV between DRHBc$^\star$ and RCHB$^\star$ for the five-inputs case.}
   \label{fig:2}
\end{figure}
To perform a sample sensitivity study of the r-process abundances to nuclear deformations (or masses) using the RCHB$^\star$ and DRHBc$^\star$ data, we select a sample regime where the mass difference between RCHB$^\star$ and DRHBc$^\star$ is significant. 
Figure~\ref{fig:2} summarizes the mass difference between RCHB$^\star$ and DRHBc$^\star$ near Germanium isotopes. 
The filled square indicates a significant discrepancy between them, exceeding 5 MeV, while open square represents a minor difference below 5 MeV variance. The maximum discrepancy in the filled square region  is $6.3$~MeV from $^{93}$As which is a deformed nucleus with  $\beta_2= 0.316$ in FRDM(2012)~\cite{Moller:2015fba}. 
From a preliminary result of DRHBc, the $\beta_2$ of $^{93}$As is -0.254~\cite{PGuo}.
In Table \ref{tab3} we list the isotopes used in our sample sensitivity study. As expected they are all deformed.
\begin{table*}[h]
  \begin{center}
  \begin{tabular}{|c|cc||c|cc|}\hline 
           \phantom{aa}  \phantom{aa}& \phantom{aa}  AME2020 \phantom{aa} & \phantom{aa}   FRDM(2012) \phantom{aa}& \phantom{aa}  \phantom{aa}& \phantom{aa}  AME2020 \phantom{aa} & \phantom{aa}   FRDM(2012) \phantom{aa} \\
           \hline
           $\ce{^{89} Ga}$ & $-$      & $0.206$ & $\ce{^{93} As}$ & $-$      & $0.316$\\     
           $\ce{^{89} Ge}$ & $0.244$  & $0.207$ & $\ce{^{94} As}$ & $-$      & $0.316$\\ 
           $\ce{^{90} Ge}$ & $-0.255$ & $0.207$ & $\ce{^{95} As}$ & $-$      & $0.328$\\        
           $\ce{^{91} Ge}$ & $-0.250$ & $0.240$ & $\ce{^{96} As}$ & $-$      & $0.329$\\        
           $\ce{^{92} Ge}$ & $-0.246$ & $0.316$ & $\ce{^{94} Se}$ & $-0.260$ & $0.339$ \\          
           $\ce{^{93} Ge}$ & $-0.238$ & $0.327$ & $\ce{^{95} Se}$ & $-0.254$ & $0.328$ \\        
           $\ce{^{94} Ge}$ & $-0.231$ & $0.327$ &  $\ce{^{96} Se}$ & $-0.252$ & $0.340$ \\          
           $\ce{^{91} As}$ & $-$      & $0.208$ & $\ce{^{97} Se}$ & $-0.252$ & $0.340$ \\           
           $\ce{^{92} As}$ & $-$      & $ 0.208$& & &\\      
           \hline
  \end{tabular}
  \end{center}
  \caption{The isotopes used in our sample sensitivity study. The values of the quadrupole deformation $\beta_2$ are taken from AME2020~\cite{Wang:2021xhn} and from  FRDM(2012)~\cite{Moller:2015fba}.}
  \label{tab3}
\end{table*}
The study of mass sensitivity will be conducted using the nuclei marked with filled squares in Table~\ref{tab3} for the magnetohydrodynamic (MHD) and collapsar jet scenarios.  Among several candidate sites for the r-process nucleosynthesis, neutron star mergers contribute to the solar r-abundances only in recent epochs, whereas MHD and collapsar jets contribute over the entire history of cosmic evolution~\cite{Kobayashi:2020jes, Yamazaki:2021yqh}. 

We finally remark that the 5 MeV difference is a bit larger than the mass variation 
in a sensitivity study of r-process abundances to nuclear masses: for example $\pm 0.5$ MeV (or $1$ MeV difference) in Ref.~\cite{Mumpower:2015hva}
and $\pm 1$ MeV (or $2$ MeV difference) in Ref.~\cite{Hao:2023bzz}.

\section{Sample Mass Sensitivity Study}
The r-process can be understood through a network of various reactions, mainly involving neutron capture, photodissociation (the inverse of neutron capture), and beta decay. Different mass tables yield varying Q-values for reactions, which are crucial for determining reaction rates. Because the Q-value represents the energy associated with a reaction, in general an increase in the Q-value leads to  a decrease in the reaction rate. These changes in reaction rates influence the equilibrium between neutron capture and photo-dissociation processes.
Fission recycling is also important in the r-process nucleosynthesis.
Some of the heavy nuclei produced in the r-process can become so massive that they undergo fission, splitting into lighter nuclei. This fission can also release a significant number of neutrons and then the released neutrons can be captured by nearby nuclei, inducing further nucleosynthesis~\cite{He2024}.

Now, we perform a sample sensitivity study of the r-process abundances to nuclear deformations (or masses) using the isotopes summarized in Table~\ref{tab3} from the RCHB$^\star$ and DRHBc$^\star$ data. For this, we use the nuclear reaction network calculation code, Libnucnet~\cite{Libnucnet} and take the thermonuclear reaction rate from JINA-REACLIB  database~\cite{JINA:REACLIB}. In addition to the masses observed in experiments, a theoretical mass model, FRDM (2012)~\cite{Moller:2015fba}, is used.  
The code has been slightly updated by adding several new reaction rates and by adopting latest experimental measurements if available~\cite{K2}.

Once the mass of a nucleus changes, the corresponding $\beta$-decay rates and neutron capture rates can also change.
The neutron capture rates are calculated with the publicly available statistical model code $\textmd{TALYS}$ \cite{Koning:2023ixl}.
We evaluate the corresponding $\beta$-decay rates by using the empirical $\beta$-decay rates developed in Ref.~\cite{zhou:2023beta}:
\begin{equation}
  \log{t_{1/2}} = c_1 - c_2 \log{Q},  
\end{equation}
where $c_1$ and $c_2$ take the values of (3.69, 4.80), (4.59, 5.08), (4.98, 5.53), and (6.32, 6.34) for even-even, even-odd, odd-even, and odd-odd parents nuclei, respectively.

For the mass sensitivity study, we consider the MHD jet supernova model and collapsar jets.
\begin{center}
{\bf MHD jet model}
\end{center}
In the MHD jet model~\cite{Nishimura:2005nz, Winteler:2012hu, Nishimura:2015nca}, rapid rotation
and strong magnetic fields can produce neutron-rich jets in polar direction along the rotational axis of the core collapse supernova, providing a favorable environment for the $r$-process.
In our simulations, the twenty-three trajectories from the MHD jet model in Ref.~\cite{Shibagaki:2015fga} are used.
\begin{figure}[ht!]
  \begin{center}
    \includegraphics[width=8cm]{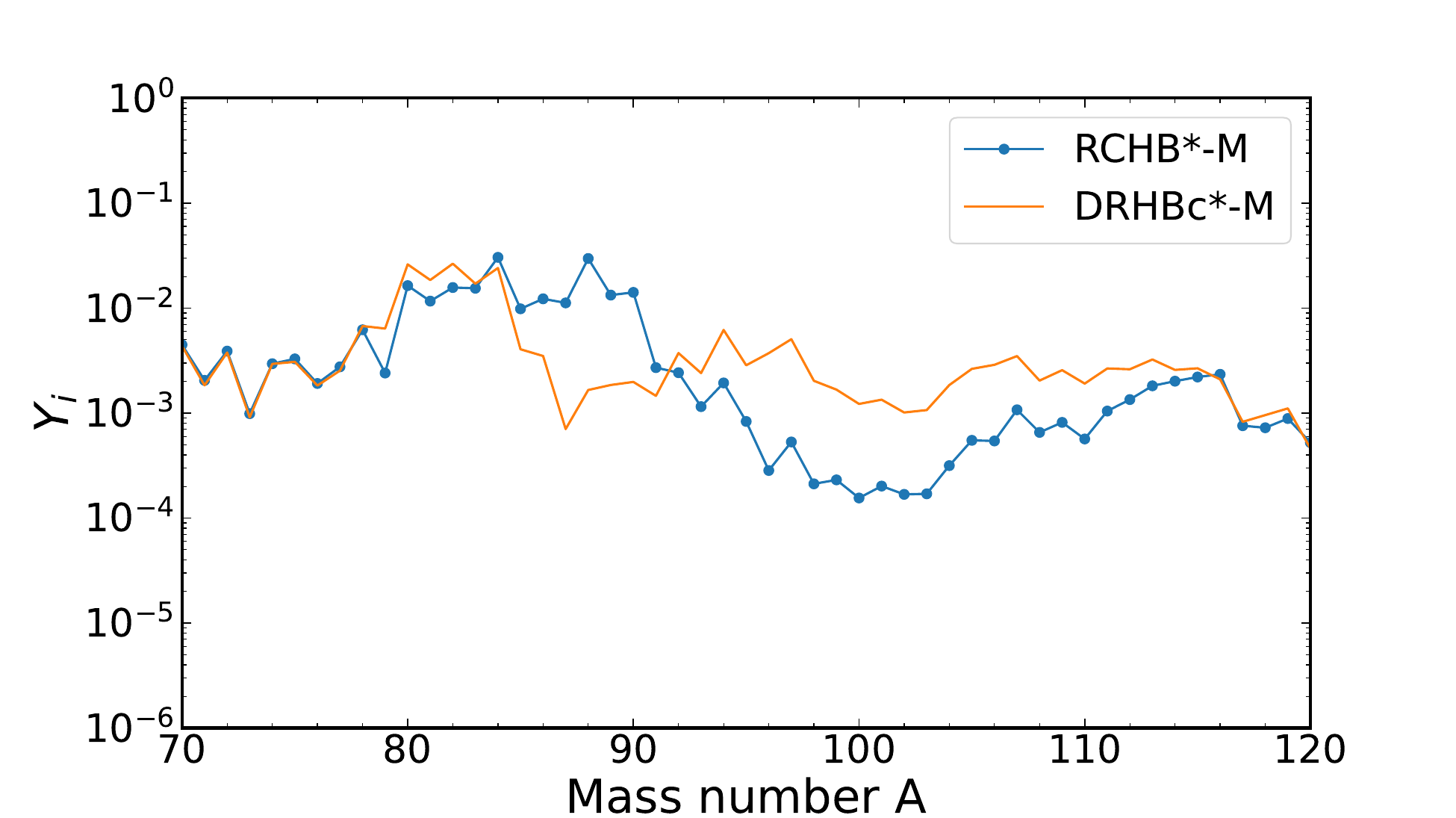}
    \includegraphics[width=8cm]{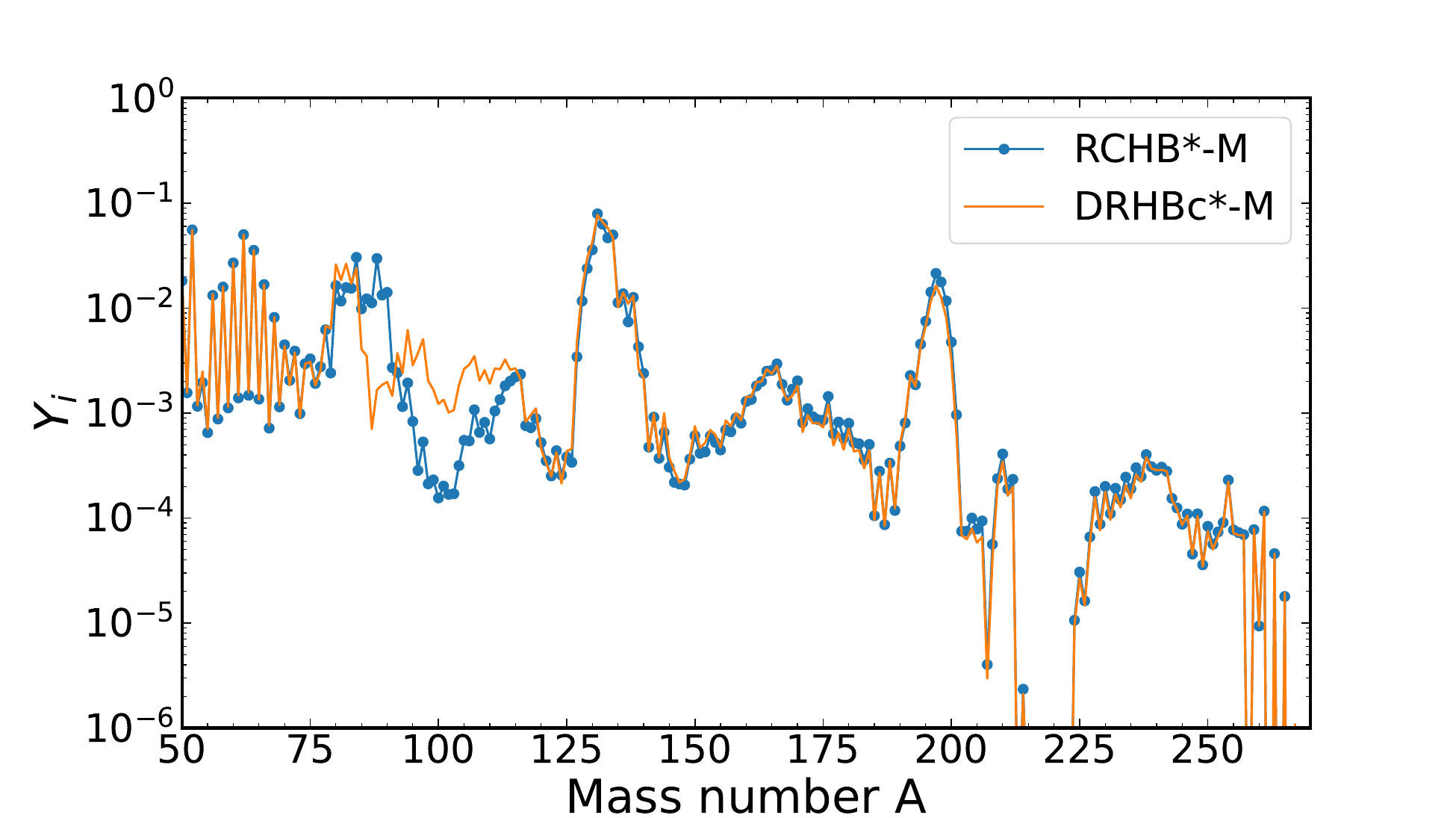}
    \includegraphics[width=8cm]{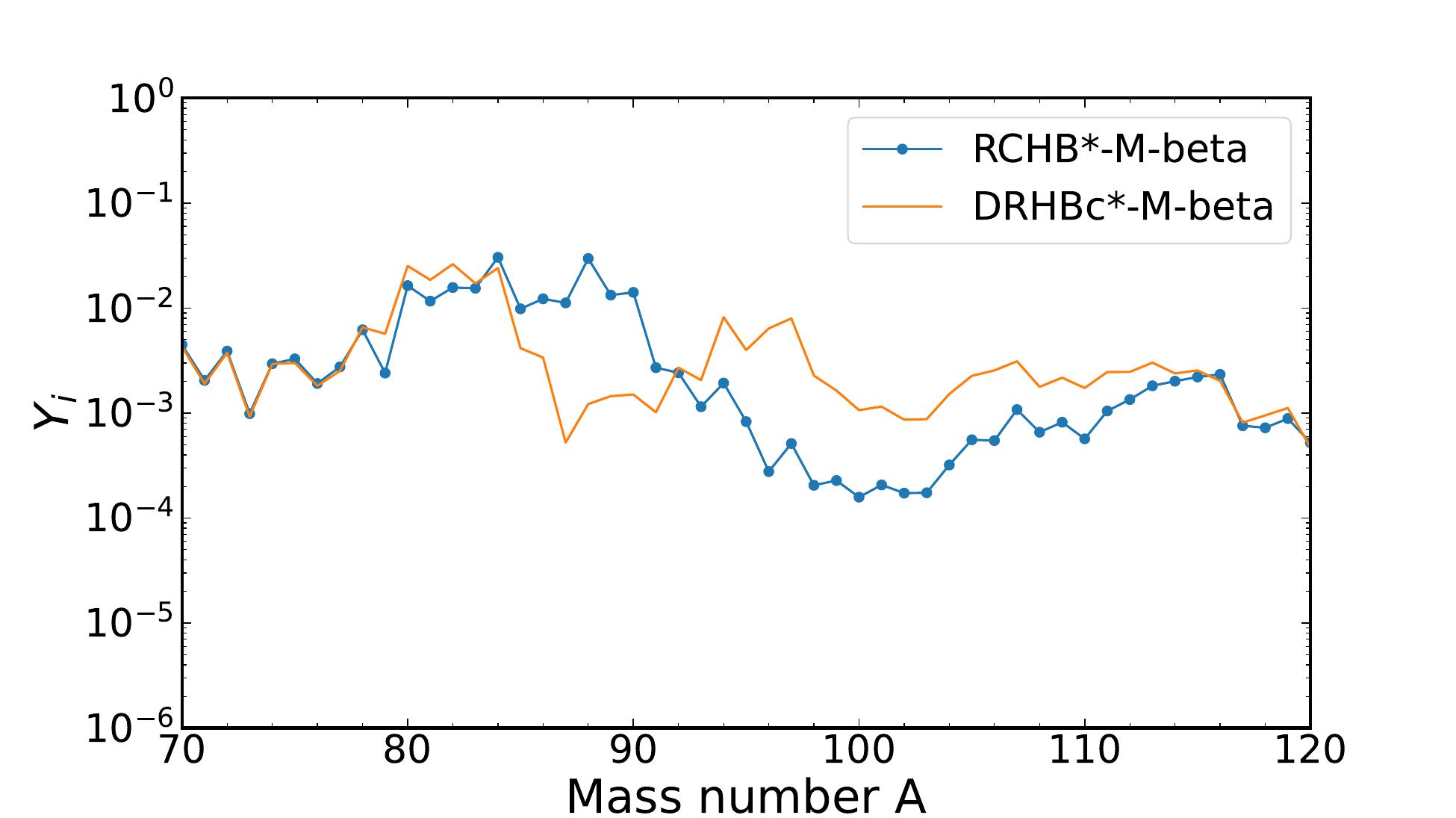}
    \includegraphics[width=8cm]{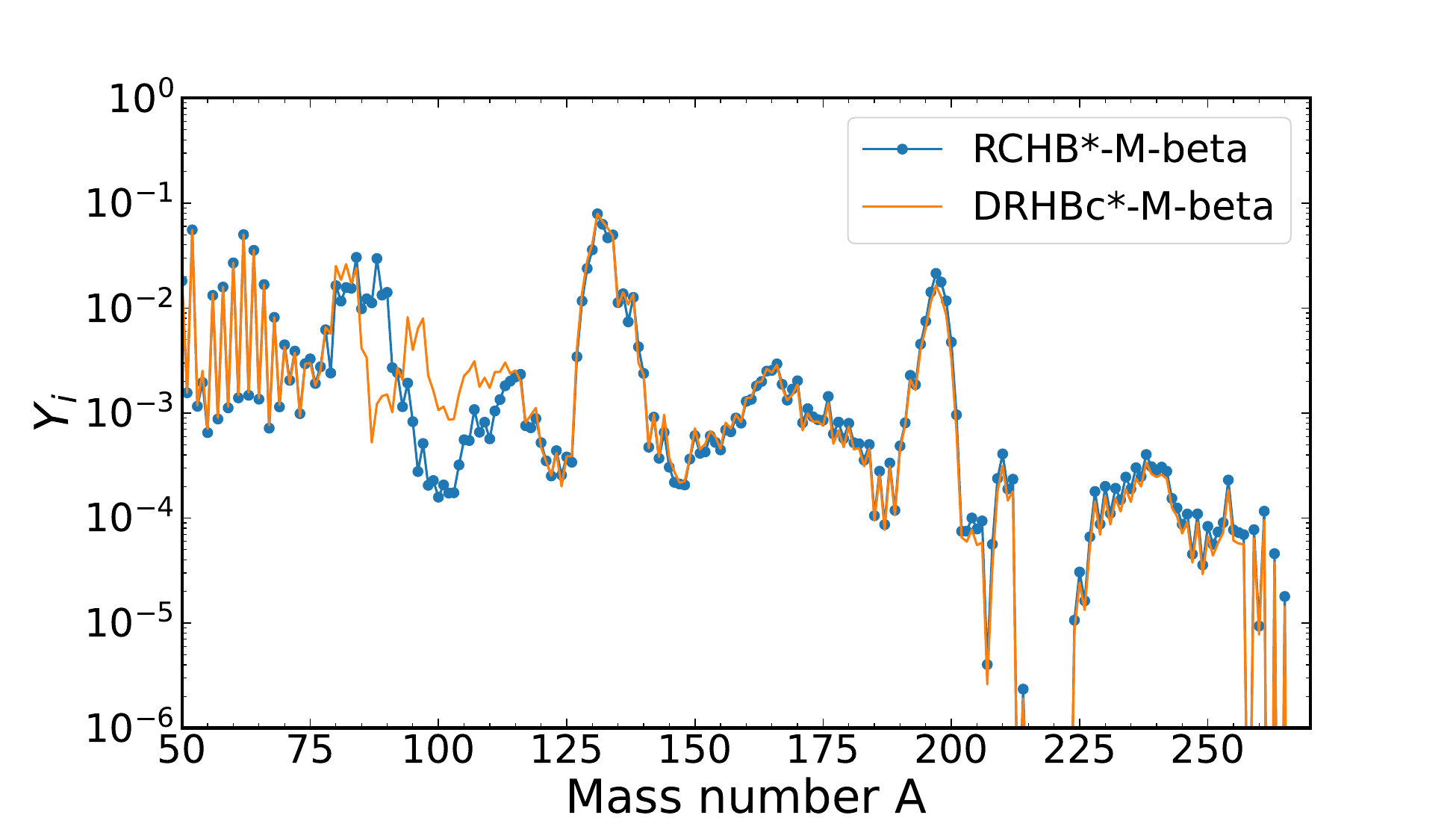}
    \includegraphics[width=8cm]{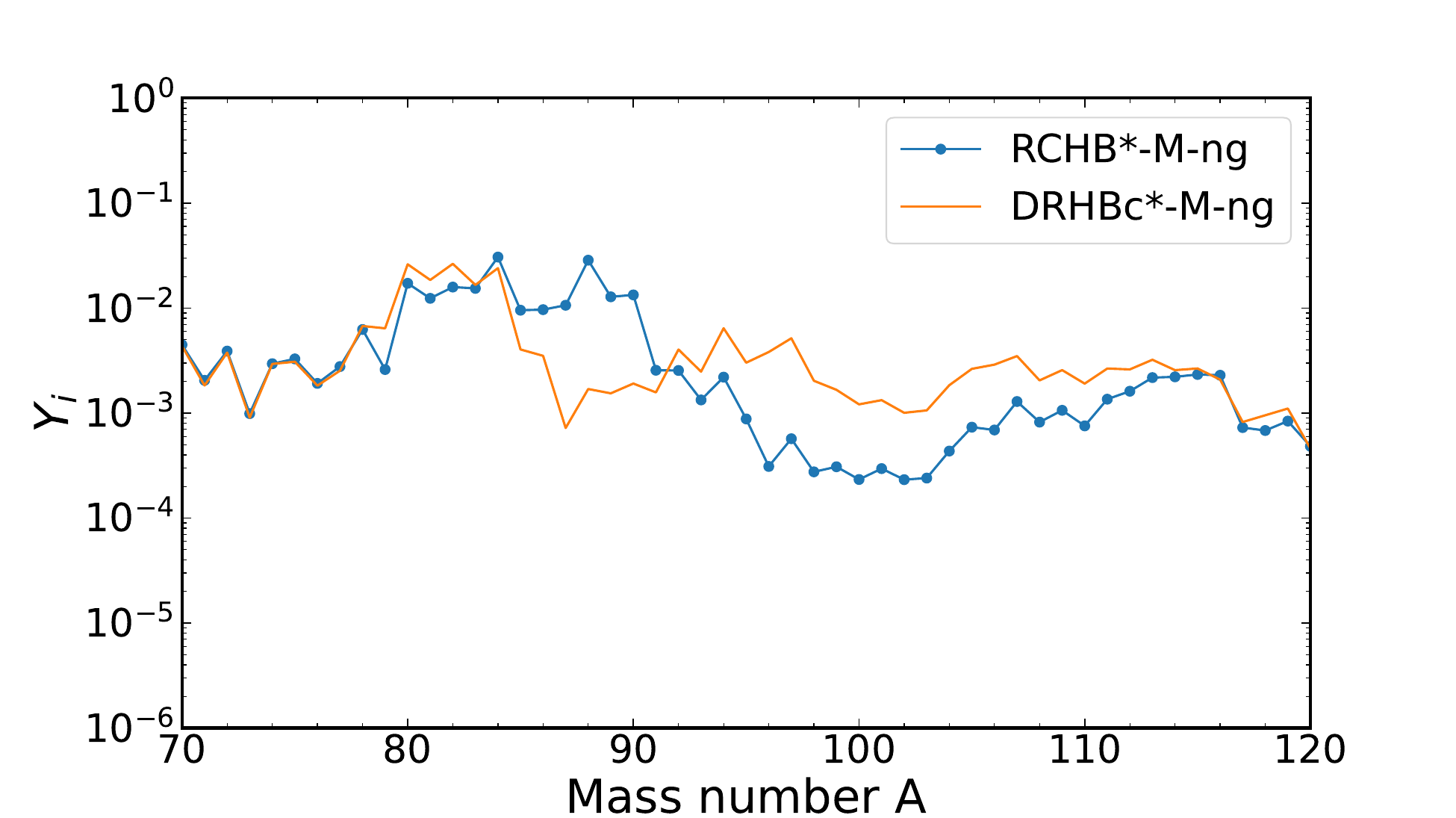}
    \includegraphics[width=8cm]{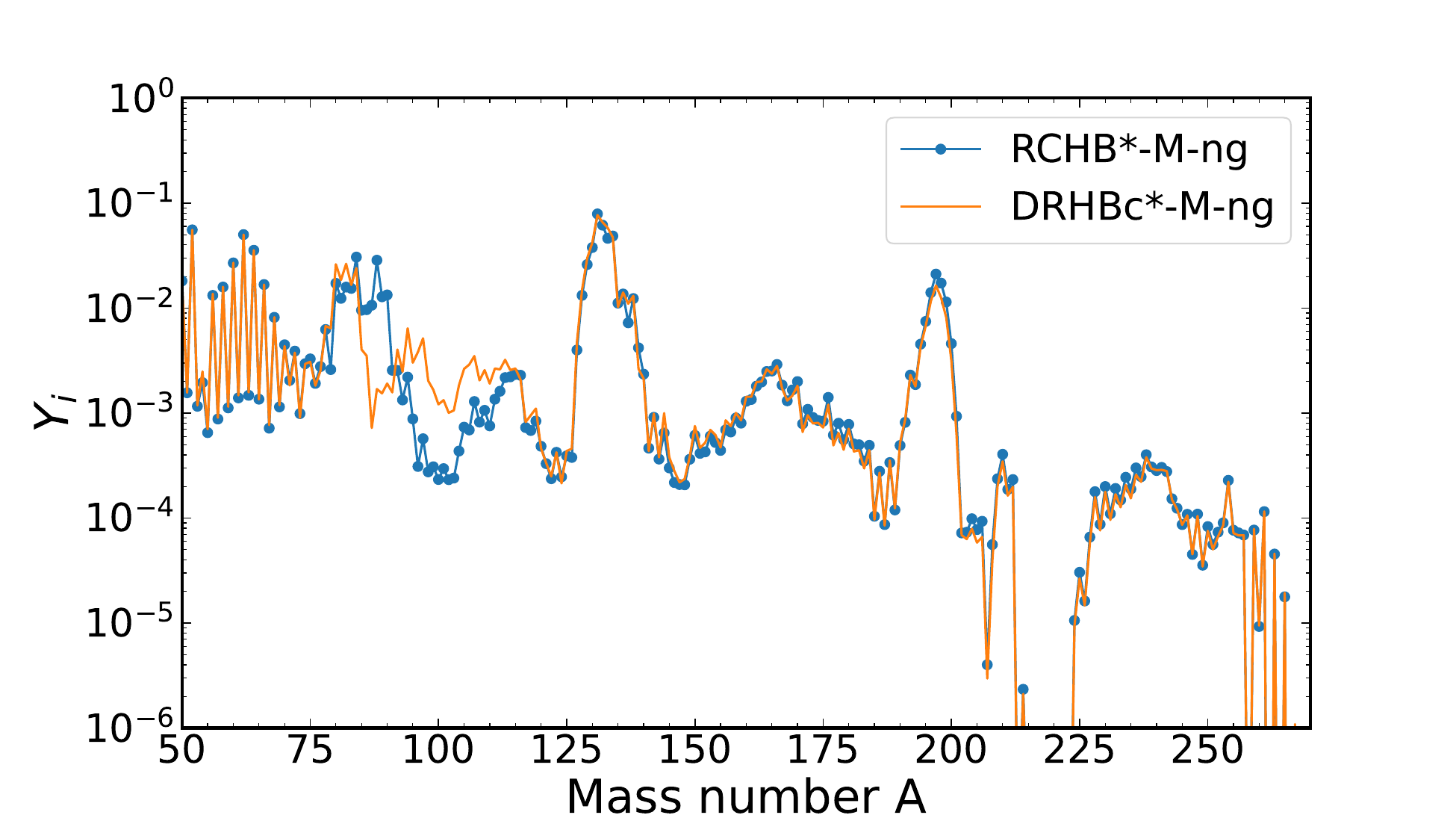}
    \includegraphics[width=8cm]{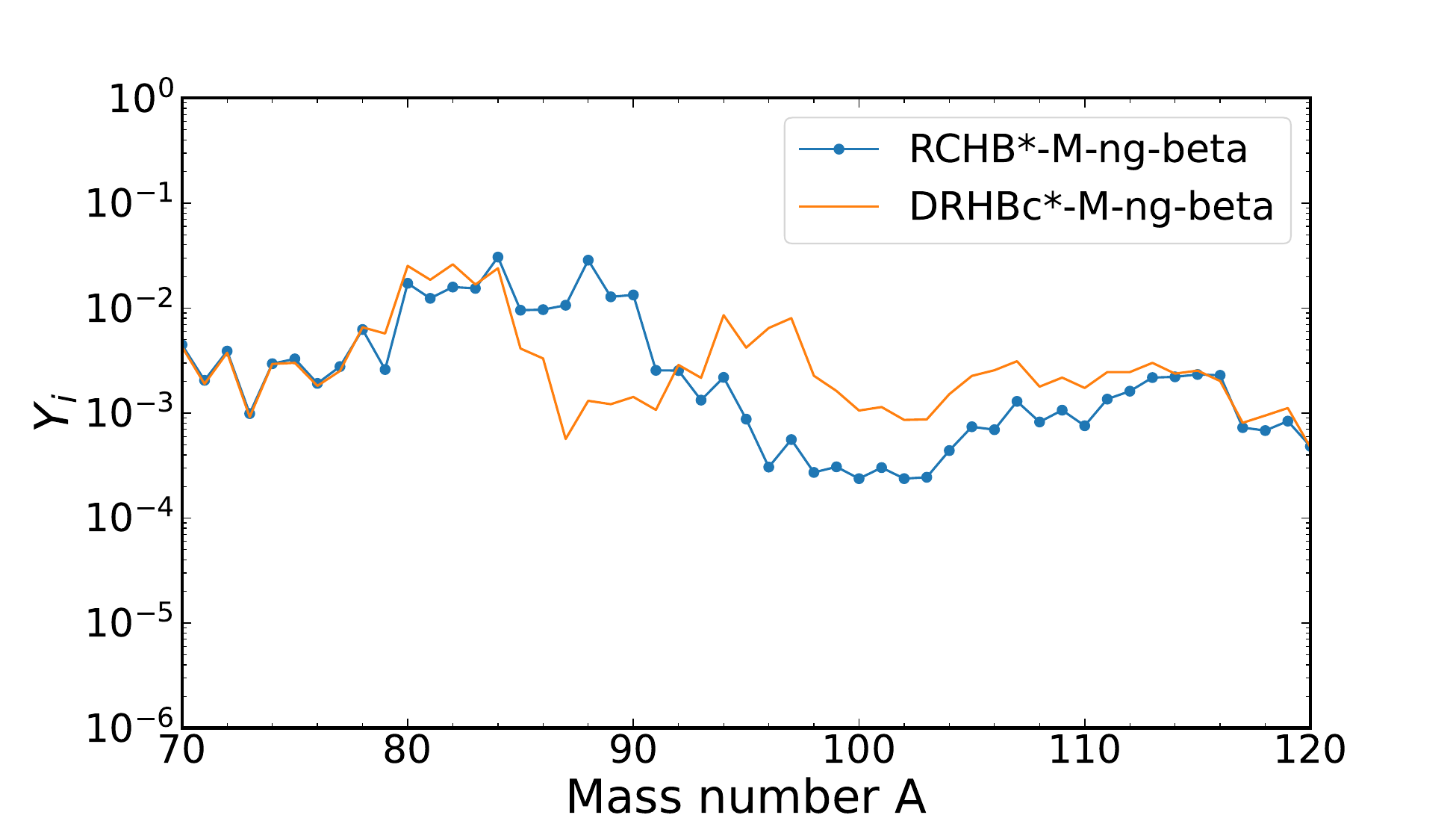}
    \includegraphics[width=8cm]{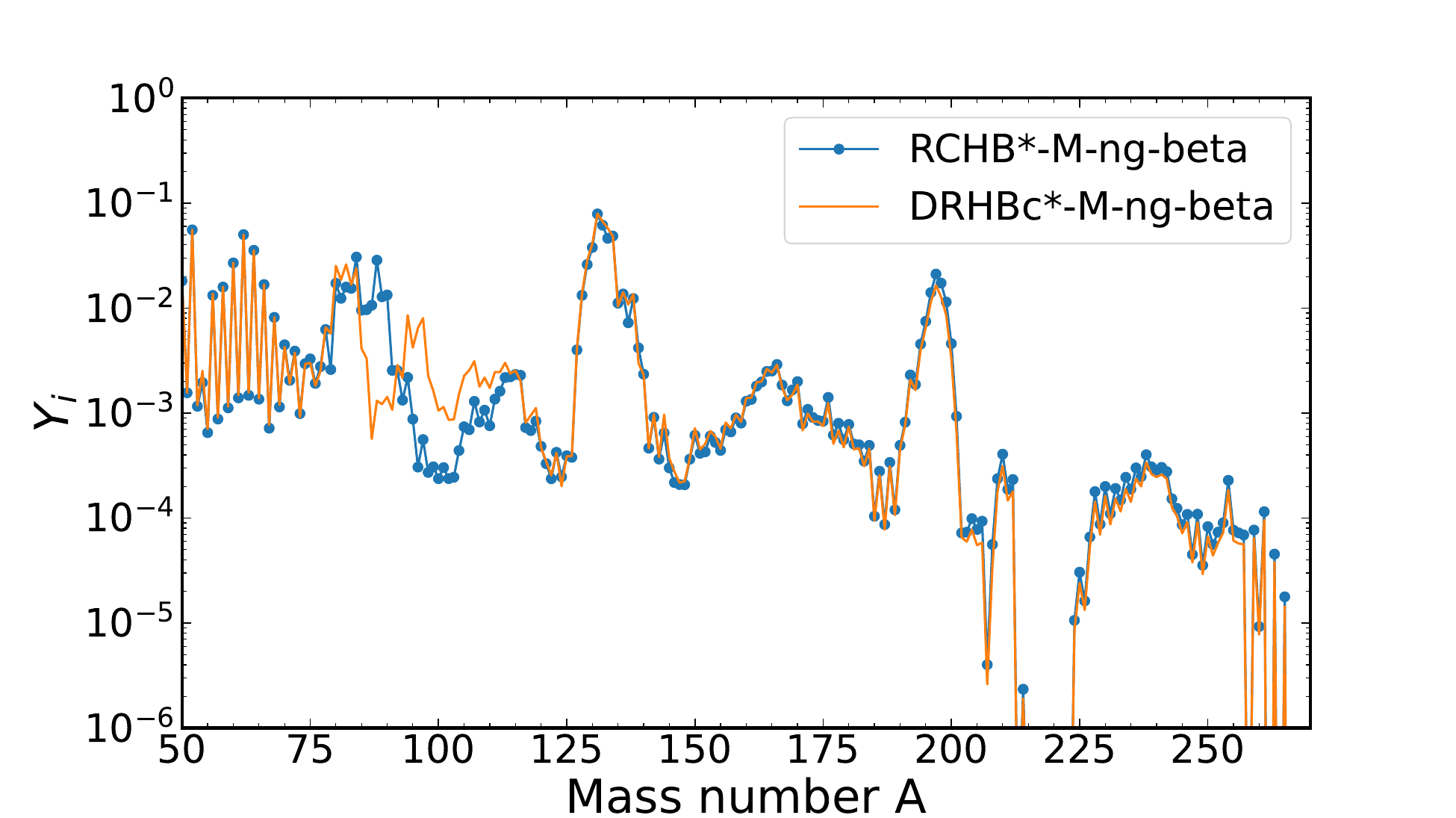}
  \end{center}
  \caption{MHD results with fission recycling: The top panel shows the effect of changing only the mass, the second panel adjusts both the mass and beta decay rate, the third panel modifies the mass and neutron capture rate, and the bottom panel varies the mass, beta decay rate, and neutron capture rate. The figures on the left are displayed on an expanded scale.}
   \label{fig:3}
\end{figure}

Figure~\ref{fig:3} shows the final r-process yield ($Y_i$) patterns obtained from the MHD jet environments with fission recycling with two different nuclear mass models: RCHB$^\star$ (blue line with dots) and DRHBc$^\star$ (orange line). 
All yield values are normalized such that $\sum Y_i = 1$, ensuring that the total abundance across all mass numbers remains consistent for comparison between different models. 
In the figure, we examine three cases: changing only the mass (top panel), altering both the mass and the beta-decay rate (second panel), and modifying the mass alongside the neutron capture rate (third panel). The bottom panel presents a case where all three —mass, beta-decay rate, and neutron capture rate— are changed simultaneously. The figures on the left use an expanded scale. It can be seen from Fig.~\ref{fig:3} that the r-process yields of RCHB$^\star$ and DRHBc$^\star$ can differ by up to two orders of magnitude in the regime from $A = 80$ to $120$.
We note that fission recycling plays a minimal role in the MHD model, as the yields of nuclei with $A > 265$ prior to fission recycling are negligible. As a result, in the MHD scenario, the final r-process yield patterns are practically identical with and without fission recycling.

Another notable feature in Fig.~\ref{fig:3} is that the lines ($Y_i$s) are intersect near $A=91\, {\rm or}\, 92$.
This behavior can be explained by examining the Q-values of neutron capture reactions using the RCHB$^\star$ and DRHBc$^\star$ mass tables. 
The differences between the Q-values from  the RCHB$^\star$ and DRHBc$^\star$ mass tables  change sign from negative to positive  as the neutron number increases.  For example, the Q-values for the $^{90}$As(n, $\gamma$)$^{91}$As reaction are 1.04 MeV for the RCHB$^\star$ mass table and 6.22 MeV for the DRHBc$^\star$ mass table. 
In contrast, the Q-values for $^{93}$As(n, $\gamma$)$^{94}$As are 1.96 MeV (RCHB$^\star$) and 1.88 MeV (DRHBc$^\star$), while for $^{96}$As(n, $\gamma$)$^{97}$As are 4.41 MeV (RCHB$^\star$) and -1.13 MeV DRHBc$^\star$. 
This indicates that neutron capture reactions involving more neutron-rich nuclei are more favorable with the DRHBc$^\star$ mass table. 
Within the mass range we explored, the DRHBc$^\star$ mass table predicts the production of more neutron-rich nuclei, leading to higher abundances of elements with $A>92$.

\begin{figure}[ht!]
  \begin{center}
  \includegraphics[width=8.8cm]{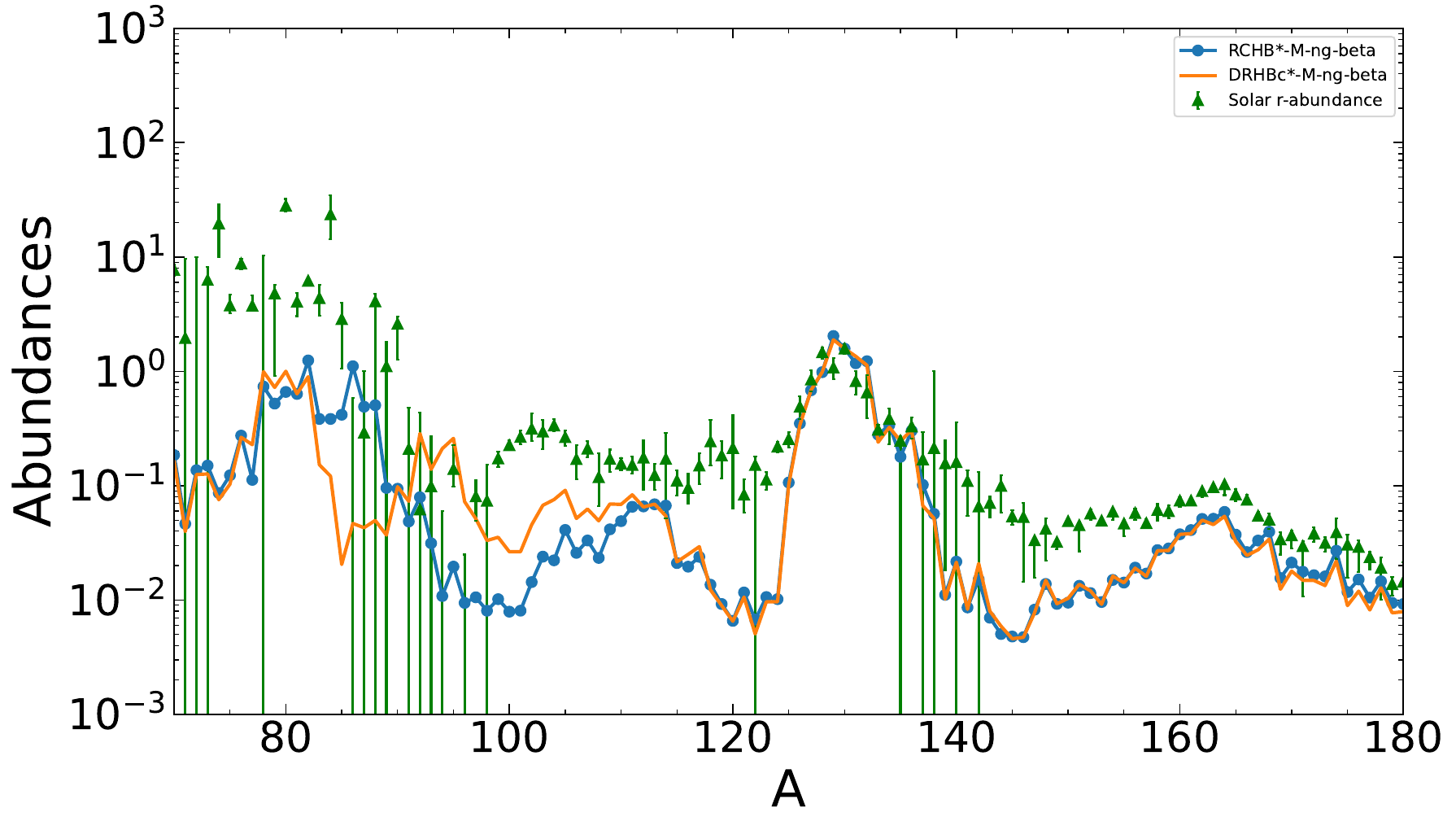}
  \end{center}
  \caption{The contribution from the MHD model with RCHB$^\star$ and DRHBc$^\star$ to the solar r-abundances.}
   \label{Solar-MHD}
\end{figure}
We finally present the contribution of the MHD model with RCHB$^\star$ and DRHBc$^\star$ to the solar r-abundances~\cite{Goriely-1999}. As shown in Fig.~\ref{Solar-MHD}, the DRHBc$^\star$ model contributes more than the RCHB$^\star$ model to the small peak near A = 104. We anticipate that the complete mass table from the DRHBc calculation will enable a more precise study of the relationship between the solar r-abundances near A = 104 and nuclear deformations.

 As it can be seen from Fig.~\ref{Solar-MHD}, there are large error bars in the solar r-abundance around A=90.  It is crucial to reduce these error bars by accurately determining the s-process components in the solar abundances, in order to better understand the role of nuclear deformations in r-process nucleosynthesis.

\begin{center}
{\bf Collapsar jet model}
\end{center}
We adopt the collapsar model developed in Ref.~\cite{Harikae:2009dz,Nakamura:2013}, where the progenitor mass is  35 $M_{\odot}$, rotational velocity is $v_{\phi}$ = 380 km/s and metallicity is approximately 0.1 $ Z_{\odot}$. In our calculations we consider  the eight representative trajectories from Ref.~\cite{Nakamura:2013}.

\begin{figure}[ht!]
  \begin{center}
  \includegraphics[width=8cm]{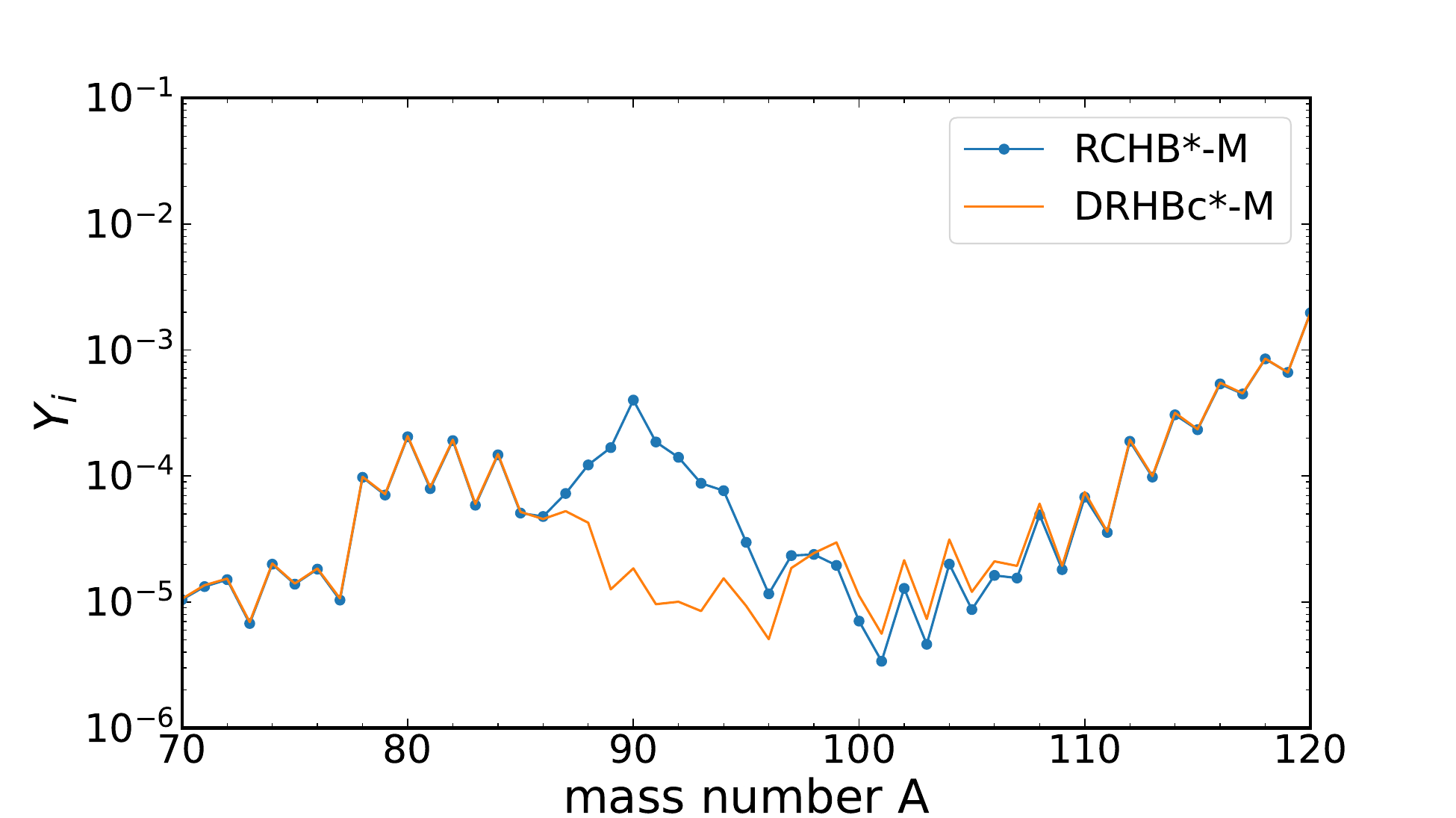}
  \includegraphics[width=8cm]{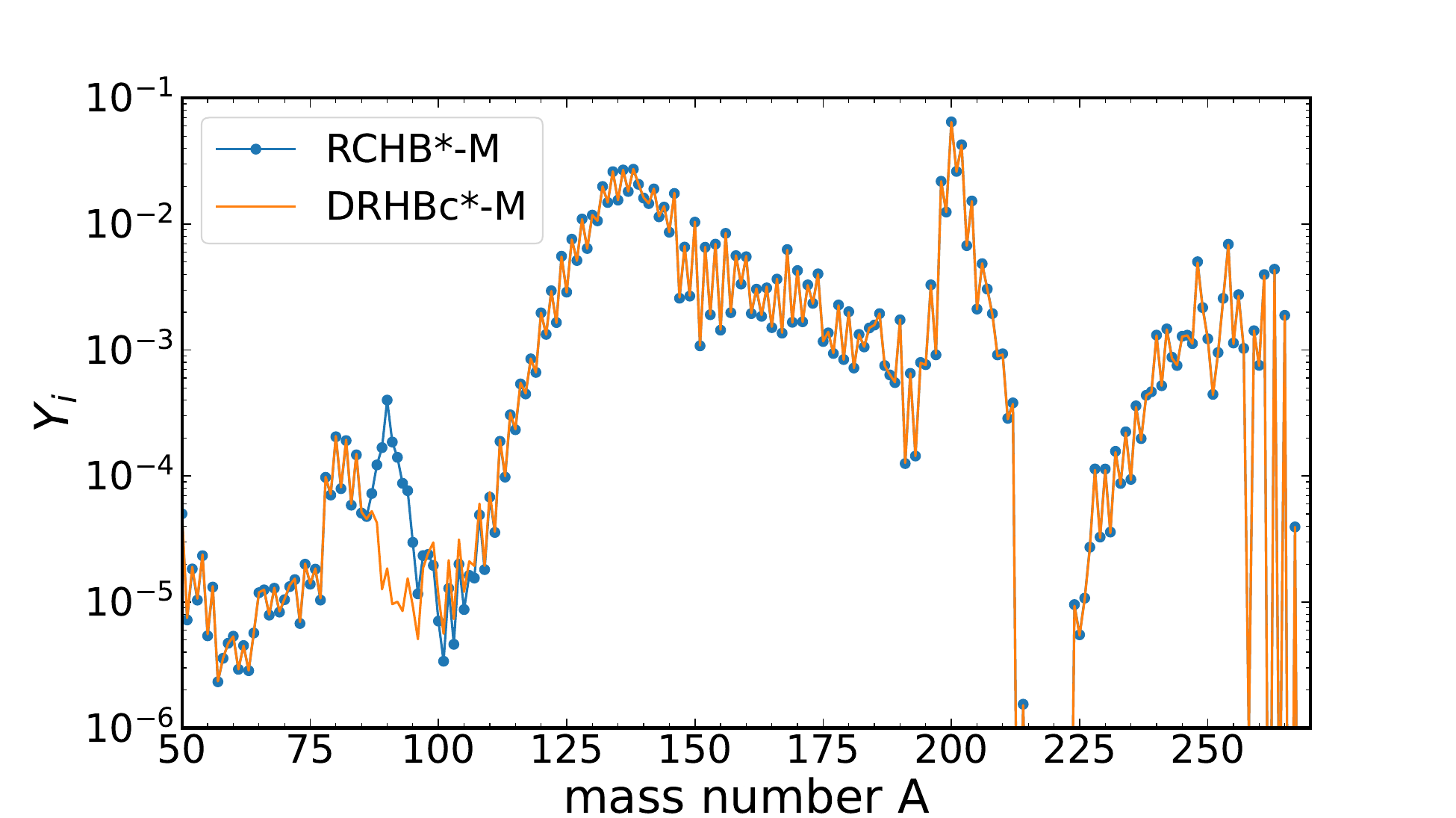}
  \includegraphics[width=8cm]{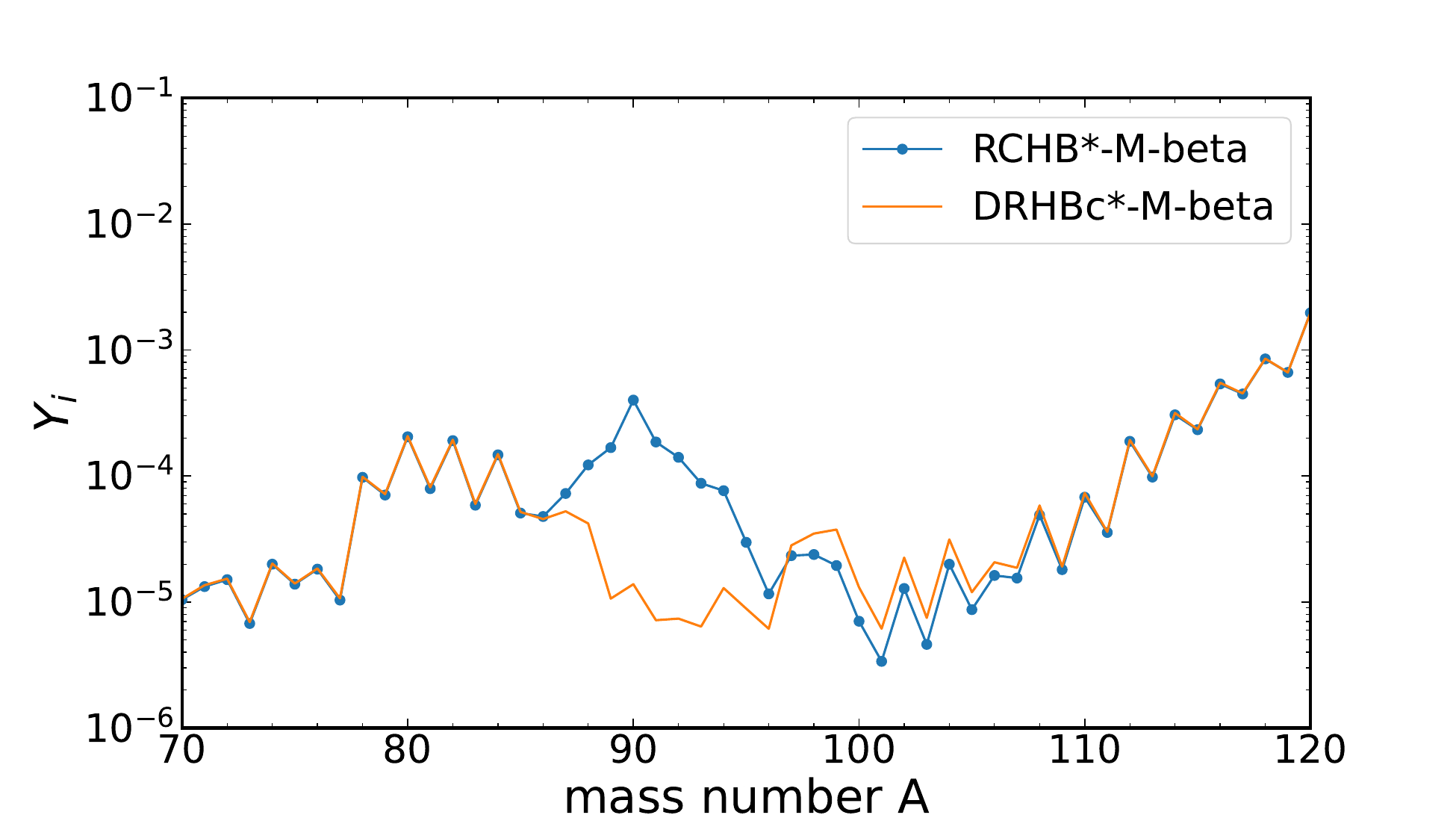}
  \includegraphics[width=8cm]{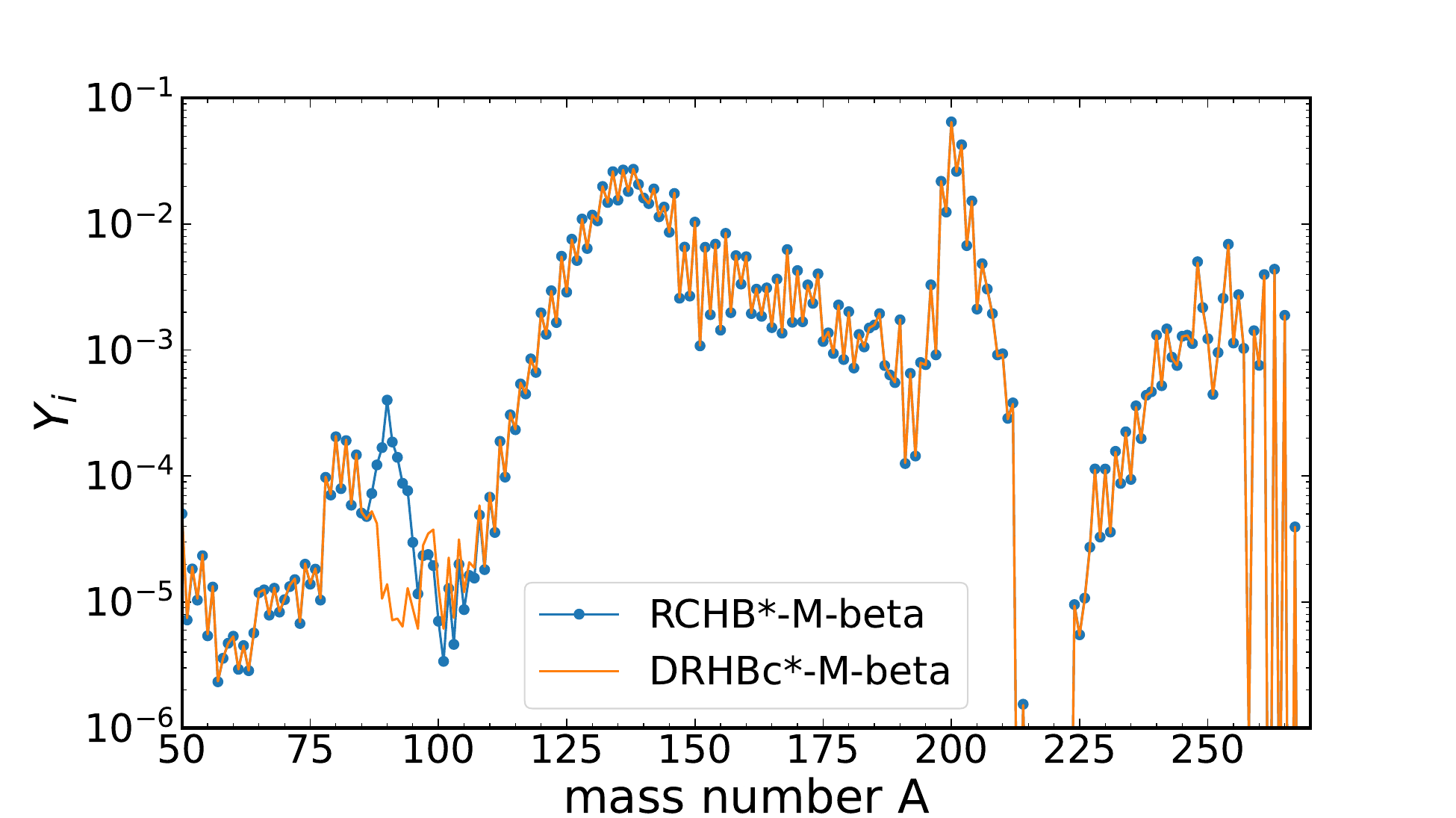}
  \includegraphics[width=8cm]{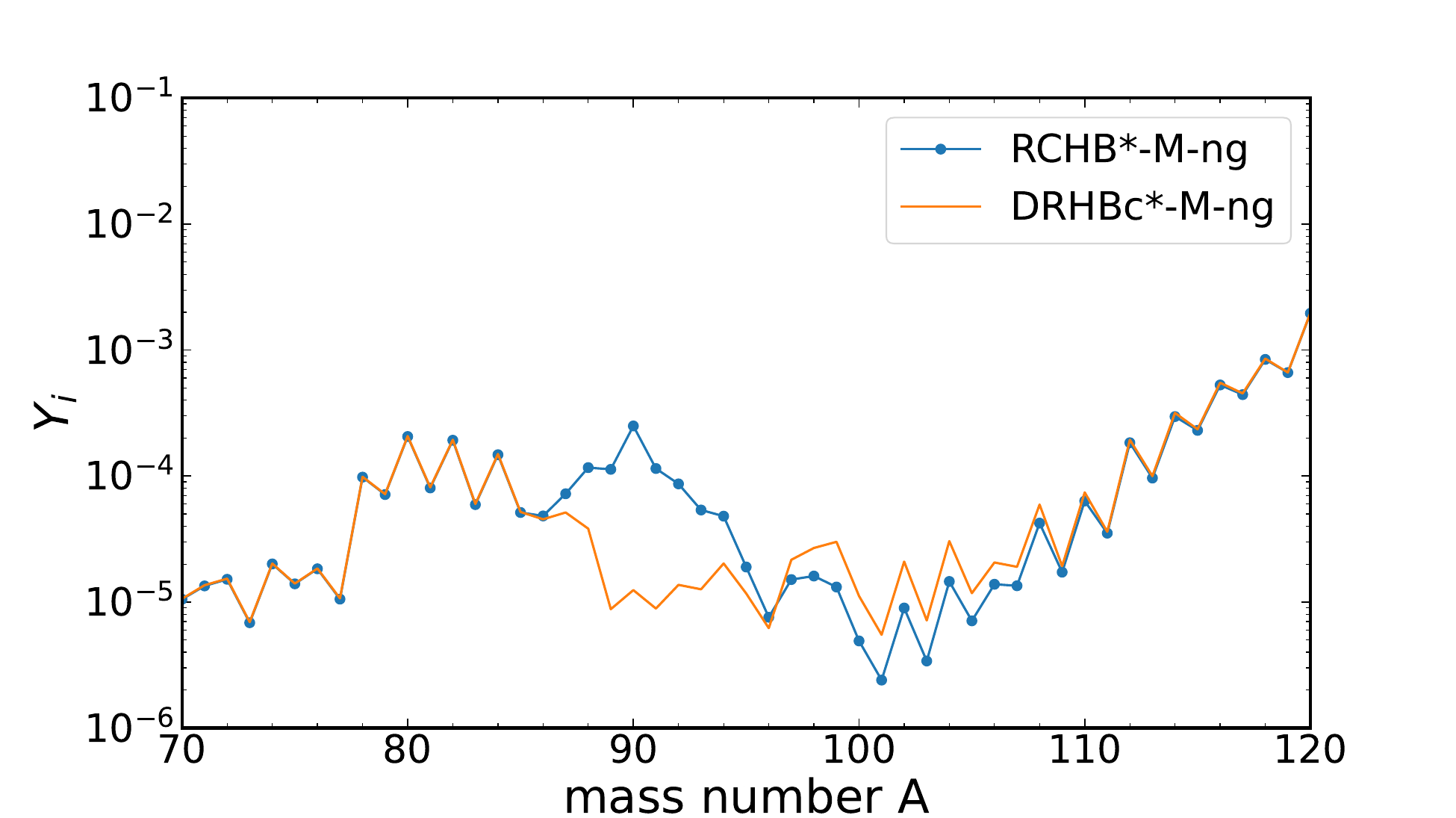}
  \includegraphics[width=8cm]{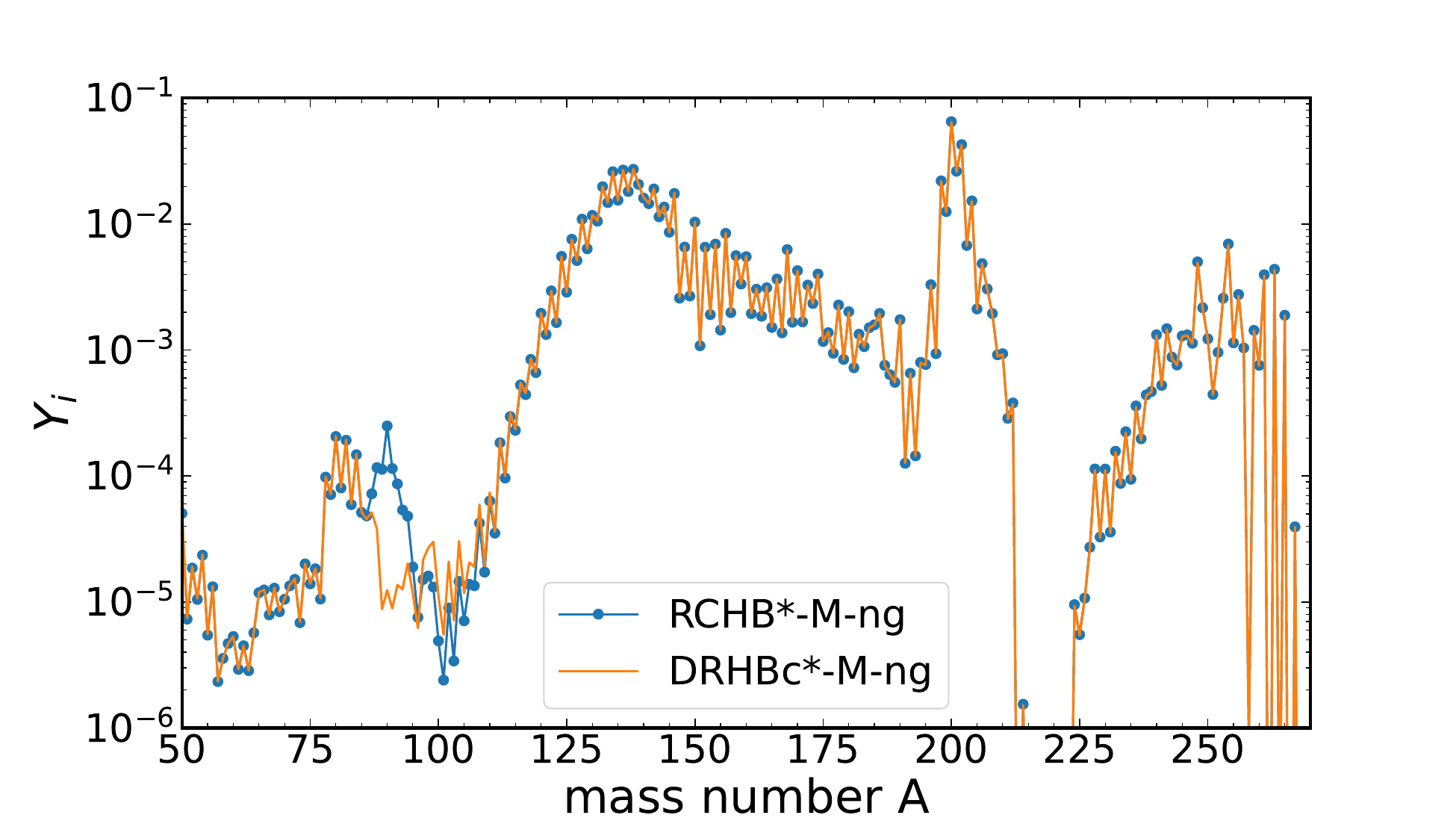}
  \includegraphics[width=8cm]{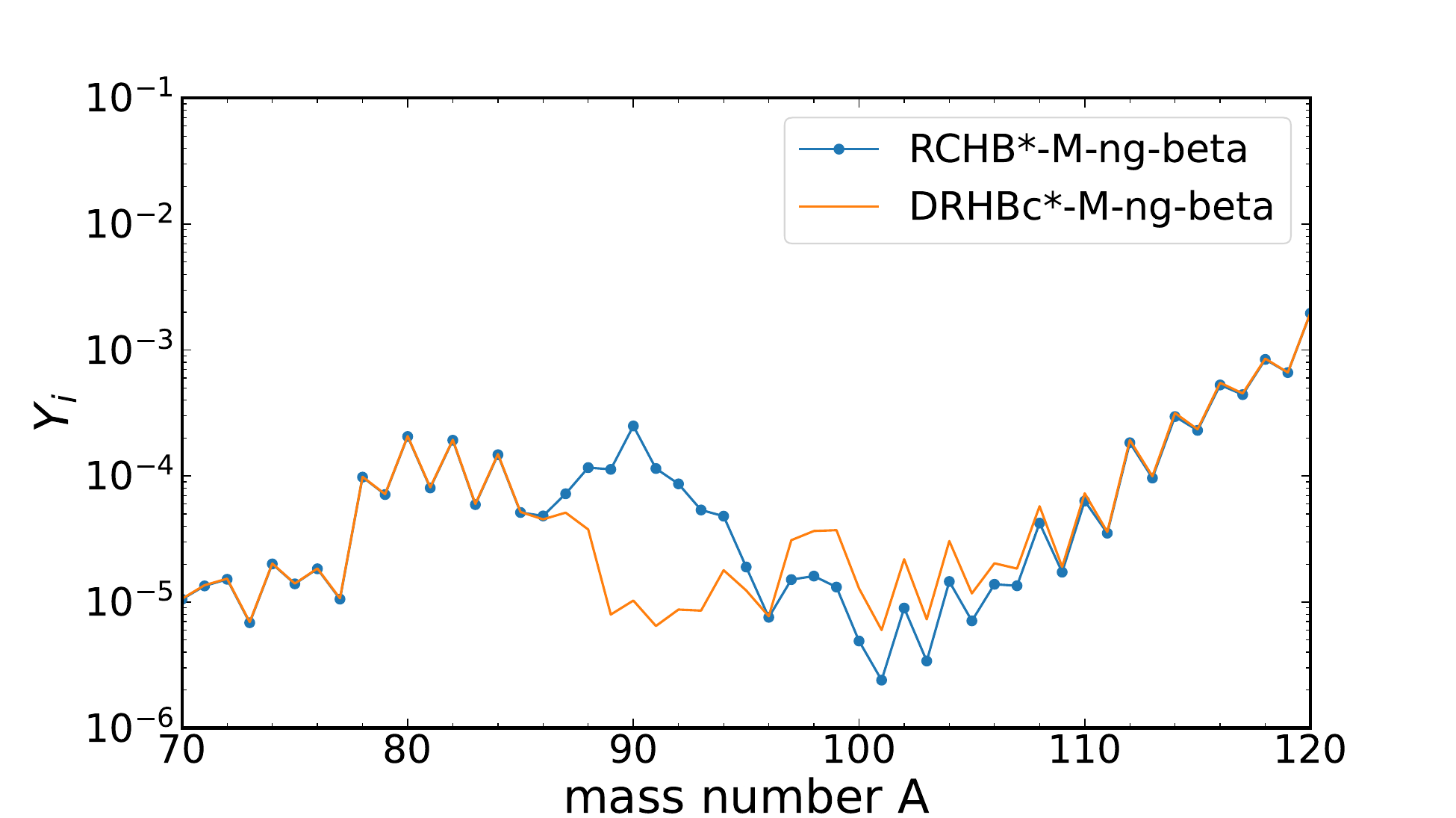}
  \includegraphics[width=8cm]{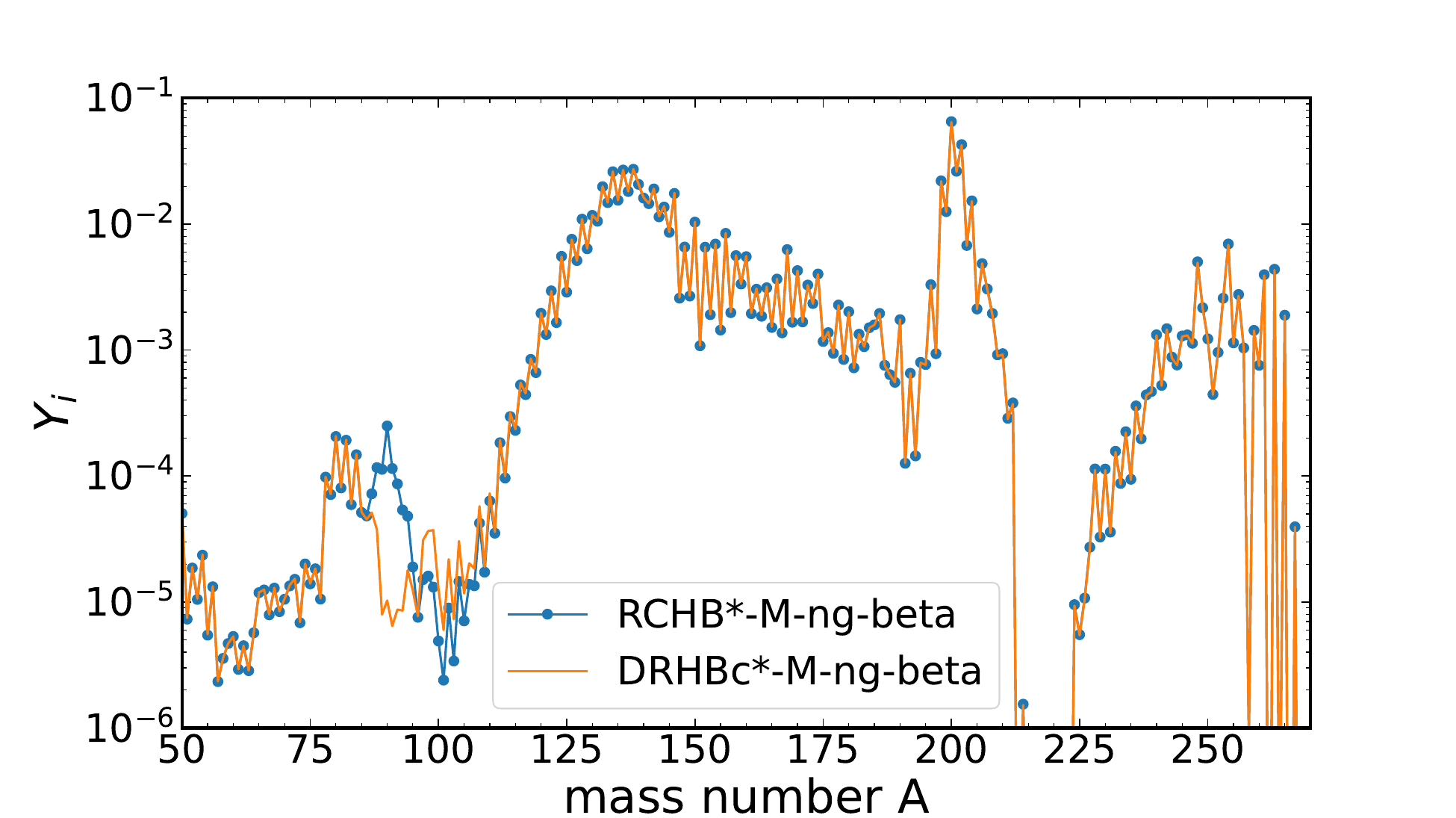}
  \end{center}
  \caption{ Collapsar results with fission recycling:  the top panel changes only the mass, the second panel modifies both the mass and beta decay rate, the third panel adjusts the mass and neutron capture rate and the bottom panel varies the mass, beta decay rate, and neutron capture rate.}
   \label{fig:5}
\end{figure}
Figure~\ref{fig:5} shows the final abundance patterns of the r-process in collapsar environments with fission recycling, using the fission rates provided in~\cite{Shibagaki:2015fga}. Unlike the MHD model scenario, fission recycling plays a crucial role in shaping the final abundance pattern in collapsar jets. In particular, the abundance distribution for nuclei at the mass region 100$<$A$<$150 is significantly redistributed by fission fragments. Therefore, the pronounced abundance differences at 100$<$A$<$120 in the MHD model (Fig.~\ref{fig:3}) are nearly eliminated in the collapsar model (Fig.~\ref{fig:5}).

This can be explained as follows: The collapsar environment is more explosive than the MHD jet environment, leading to the formation of more neutron-rich nuclei as a result of the active occurrence of neutron-capture reactions. In this context, nuclear reactions involving elements from $^{89}$Ga to $^{97}$Se contribute relatively little. As the results show, the r-process predominantly produces heavy elements by consuming lighter ones, underscoring the critical role of fission recycling in this environment.

\section{Summary and discussion}
As  an initial step in studying the role of nuclear deformations in r-process nucleosynthesis, we conducted a preliminary sensitivity study of the r-process. To complete the DRHBc mass table for nuclear binding energies, we employed the DNN method, which resulted in an improved RMS deviation of $0.842$ MeV between the AME2020 and DRHBc$^\star$ datasets; the RMS deviation between AME2020 and the even-Z DRHBc mass table is 1.433 MeV. We also compared the mass of the neutron-rich isotopes newly measured in~\cite{Xian:2024ixi} with our results in DRHBc$^\star$ and obtained the RMS deviation of 0.725 MeV. This
deviation is a bit improved compared with 0.842 MeV in Table~\ref{tab1}, which indicates that DRHBc$^\star$ (and also DRHBc)
may work better for exotic nuclei.
Additionally, we compared the one-neutron separation energies for the isotopes of Ca, Dy, U, and As between DRHBc and DRHBc$^\star$ and the AME2020 data, and observed that our results show reasonable agreement.

We then calculated the r-process abundances by using portions of the RCHB$^\star$ and DRHBc$^\star$ mass tables, where  the mass difference exceeds 5 MeV, in the astrophysical sites of the r-process:MHD  and collapsar jets. 
We found that r-process abundances in these two sites are sensitive to the mass difference between RCHB$^\star$ and DRHBc$^\star$, particularly within the mass range of $A=80-120$. Since the primary difference between the two mass tables lies in nuclear deformations, our study suggests that deformations play a significant role in r-process nucleosynthesis.
In the future, once the complete DRHBc mass table becomes available, we plan to conduct an in-depth study of the relationship between the solar r-process abundances (particularly near A=104) and nuclear deformations.

\section*{Acknowledgments}
Y.K. thanks the DRHBc collaboration members for helpful comments.
This work was supported in part by the Institute for Basic Science (IBS-R031-D1, 2013M7A1A1075764),
the National Key R$\&$D Program of China (2022YFA1602401) and the National Natural Science Foundation of China (No. 12335009 $\&$ 12435010).

\end{document}